\documentclass{aa}
\usepackage{natbib,twoopt}
\usepackage{amsmath}

\usepackage{color}

\usepackage[breaklinks=true]{hyperref} 
\makeatletter
  \newcommandtwoopt{\intheteads}[3][][]{\href{http://adsabs.harvard.edu/abs/#3}%
    {\def\hyper@linkstart##1##2{}%
     \let\hyper@linkend\@empty\citealp[#1][#2]{#3}}}
  \newcommandtwoopt{\citepads}[3][][]{\href{http://adsabs.harvard.edu/abs/#3}%
    {\def\hyper@linkstart##1##2{}%
     \let\hyper@linkend\@empty\citep[#1][#2]{#3}}}
  \newcommandtwoopt{\citetads}[3][][]{\href{http://adsabs.harvard.edu/abs/#3}%
    {\def\hyper@linkstart##1##2{}%
     \let\hyper@linkend\@empty\citet[#1][#2]{#3}}}
  \newcommandtwoopt{\citeyearads}[3][][]%
    {\href{http://adsabs.harvard.edu/abs/#3}
    {\def\hyper@linkstart##1##2{}%
     \let\hyper@linkend\@empty\citeyear[#1][#2]{#3}}}
\makeatother

\usepackage{graphicx}
\usepackage{txfonts}
\begin{document}

   \title{SISSI: Supernovae in a stratified, shearing interstellar medium}

   \subtitle{I. The geometry of supernova remnants}

   \author{Leonard E. C. Romano
          \inst{1}\fnmsep\inst{2}\fnmsep\inst{3}\fnmsep\thanks{Corresponding author: Leonard E. C. Romano\\\email{lromano@usm.lmu.de}}
          \and
          Manuel Behrendt\inst{1}\fnmsep\inst{2}
          \and
          Andreas Burkert\inst{1}\fnmsep\inst{2}\fnmsep\inst{3}
          }

   \institute{Universitäts-Sternwarte, Fakultät für Physik, Ludwig-             Maximilians-Universität München, Scheinerstr. 1, D-               81679 München, Germany
         \and
             Max-Planck-Institut für extraterrestrische Physik, Giessenbachstr. 1, D-85741 Garching, Germany
        \and
             Excellence Cluster ORIGINS, Boltzmannstr. 2, D-85748 Garching, Germany
             }

 
  \abstract
   {}
   {We introduce the SISSI (Supernovae In a Stratified, Shearing Interstellar medium) simulation suite, which aims to enable a more comprehensive understanding of supernova remnants (SNRs) evolving in a complex interstellar medium (ISM) structured under the influence of galactic rotation, gravity and turbulence.}   
   {We utilize zoom-in simulations of 30 SNRs expanding in the self-consistent ISM of a simulated isolated disk galaxy -- the first such simulations achieving sub-parsec resolution in a galactic context.
   The ISM of the galaxy is resolved down to a maximum resolution of $\sim 12\,\text{pc}$, while we achieve a zoomed-in resolution of $\sim 0.18\, \text{pc}$ in the vicinity of the explosion sources.
   We compute the time-evolution of the SNRs' geometry and compare it to the observed geometry of the Local Bubble.
   }
   {During the early stages of evolution ($\lesssim 1 \, \text{Myr}$), SNRs are well described by existing analytical models.
   SNRs depart from spherical symmetry, within $ \sim 1\, \%$ of an orbit, earlier than galactic shear alone predicts, with deformation timescales correlating strongly with local density variations.
   The minor axis of oblate SNRs preferably aligns with the galactic poles, while the major axis of prolate SNRs aligns with galactic rotation, with a pitch angle in the range of $10 - 60^{\circ}$ -- in agreement with the expectation from galactic shear, suggesting a shear-related origin, such as interactions with shear-deformed substructure. 
   A comparison with the geometry of the Local Bubble reveals that it might be slightly younger than the previously estimated $\sim 14\,\text{Myr}$, but otherwise has a standard morphology for a SNR of its age and size.
   }
   {Studying the geometry of SNRs can reveal valuable insights about the complex interactions shaping their dynamical evolution.
   Future studies targeting the geometry of Galactic SNRs may use these insights to obtain a clearer picture of the processes shaping the Galactic ISM. 
   }


   \maketitle
%

\section{Introduction}
Advances in observational techniques over the last decades have made it possible to study the three-dimensional (3D) geometry of structures in the nearby galactic interstellar medium \citep[hereafter ISM, e.g.][]{1992A&A...258..104A, 2019A&A...625A.135L, 2024A&A...685A..82E}. Of particular interest is the Local Bubble \citep[hereafter LB][]{1987ARA&A..25..303C, 2021ApJ...920...75L}, a diffuse, X-ray emitting cavity, with a diameter of several hundred parsec, which curiously we are observing right from the center \citep{2022Natur.601..334Z, 2024A&A...690A.399Y}.
The LB is believed to be a superbubble (SB) evacuated due to the collective feedback from massive stars, such as ionizing radiation \citep{2021ApJ...920...75L}, stellar winds \citep{1998ApJ...498..689H} and supernovae \citep[hereafter SNe][]{2006A&A...452L...1B, 2021Sci...372..742W}.

The geometry of SBs and supernova remnants (SNRs) provides a valuable tool for understanding phenomena such as galactic outflows, chemical enrichment and star-formation, with both observations and theory. 
Moreover, while the LB is to date the only SB whose 3D geometry has been studied in great detail, novel techniques and a wealth of data will enable the study of many more Galactic SBs \citep{2020A&A...639A.138L, 2024A&A...685A..82E}.
Despite the lack of 3D information, extragalactic observations also provide hints to the geometry of SBs \citep{2023ApJ...944L..24W, 2024ApJ...960...81J}.
In order to be able to interpret this wealth of data, predictions from numerical simulations and analytical models for the geometry of SNRs and SBs are required.

Over the last five decades, the evolution of spherical SBs expanding into a uniform ISM has been studied in great detail \citep[e.g.][]{1974ApJ...188..501C, 1988ApJ...334..252C, 1999ApJS..120..299T}.
While these efforts have provided useful intuition for the different processes dominating the dynamics of expanding SBs and shaped the theoretical methods used to describe their evolution \citep{2015ApJ...802...99K, 2024ApJ...965..168R}, they lack the complexity needed to explore the physical processes governing the departure from spherical symmetry.

The processes that might deform SNRs are manifold. 
It has been recognized early on that blastwaves expanding into a vertically stratified atmosphere are stretched out along the density gradient \citep{1960SPhD....5...46K, 1969JFM....35...53L}.
SNRs have been found to preferentially expand into low density channels, following the density structure of the ambient ISM, shaped by gravity and turbulence \citep{2015ApJ...802...99K, 2019MNRAS.485.3887O, 2023MNRAS.523.1421M, 2025MNRAS.540.1124L}.
Moreover, galactic shear might stretch out a SB along the direction of rotation \citep{1987A&A...186..287T, 1995RvMP...67..661B}.

Observations of starburst galaxies reveal that many galaxies host galactic outflows \citep{2022ApJ...933..222X}, suggesting that vertical stratification plays an important role in shaping the geometry of SBs, provided they are powered by a sufficiently strong source.
Studies of SBs in nearby star-forming galaxies report ellipsoidal geometries, aligned with the galactic rotation \citep{2023ApJ...944L..24W}, suggesting that galactic shear might be at play.
However, from the same observations it becomes clear that density structures, such as low density channels and high-density filaments align themselves in the same way \citep{2024ApJ...975...39X}, making it difficult to disentangle the role of shear and density structure in shaping the geometry of SNRs.

While these studies, have shown the effectiveness of these various physical processes in deforming SNRs in isolation, there is only little work, addressing how they affect the geometry in concert \citep[e.g.][who however neglect radiative cooling]{2024ApJ...960...81J}.
Indeed, most studies investigating the effect of stellar feedback in turbulent, stratified, and occasionally shearing media, focus on the collective effect stellar feedback has on the average properties of the multi-phase ISM and galactic outflows \citep[e.g.][]{2005A&A...436..585D, 2015MNRAS.454..238W, 2018MNRAS.481.3325F, 2017ApJ...846..133K}.
However, a clear picture of how the different processes affecting SNR geometry compete remains unavailable.

In this paper, we present the SISSI (Supernovae In a Stratified, Shearing ISM) simulation suite, which aims to address this gap and enable a more comprehensive study of the phenomenology of SNRs.
The SISSI project, which aims to evolve well resolved SNRs in a realistic, but controlled environment, will enhance our theoretical understanding of the complex interaction of SNRs with their environment and provide future observational studies with new tools for disentangling the complex physics of SNRs in the galactic ISM. 

The remainder of this paper is organized as follows. In Sects. \ref{sec:numerics} and \ref{sec:analysis} we describe the numerical and analysis methods and give a description of the SISSI simulation suite.
In Sects. \ref{sec:time_evolution} and \ref{sec:geometry} we give an overview of the time evolution of our simulated sample of SNRs as well as an analysis of the geometry.
We discuss our results in Sect. \ref{sec:discussion}. 
Finally, we summarize our findings and conclude in Sect. \ref{sec:conclusion}. 
In the Appendix we present the properties of the ISM of our simulated galaxy, and provide some additional background to some of models and data used in our analysis.

\section{Numerical methods}\label{sec:numerics}
We model the evolution of SNRs embedded in an isolated disk galaxy, using the adaptive mesh refinement (AMR) code {\sc ramses} \citep{2002A&A...385..337T}, which solves the system of hydrodynamic equations on a finite volume, cartesian grid using a second-order unsplit Godunov method (MUSCL scheme). 
The code reconstructs variables at the cell interfaces from the cell-centered values utilizing the HLLC Riemann solver with MinMod total variation diminishing scheme \citep{1994ShWav...4...25T}.
{\sc ramses} employs a conjugate gradient method and cloud-in-cell interpolation of particle contributions to solve the Poisson equation. 

We relate the gas pressure and internal energy using an adiabatic index of $\gamma = 5/3$.
We implement radiative cooling and heating based on the {\sc HEIKOU} integration scheme (M. Behrendt et al. 2025, in prep.), which is based on the exact integration scheme \citep{2009ApJS..181..391T, 2017MNRAS.470.1017Z}, utilizing the UVB\_dust1\_CR0\_G0\_shield0 cooling table from \citet{2020MNRAS.497.4857P} at solar metallicity.
We model star-formation by allowing gas with densities $n_{\text{H}} > 100 \, \text{cm}^{-3}$ and temperatures $T < 150 \, \text{K}$ to form star particles with $m_{\star} = 10^3\, \text{M}_{\odot}$ at a rate given by a local Schmidt-law \citep[see e.g.][]{1992ApJ...391..502K, 2003MNRAS.339..289S, 2019MNRAS.484.2632S, 2022ApJS..262....9O}, with $\epsilon_{\text{ff}} = 1 \, \%$.

The simulation is separated into two stages.
In the first stage, we relax an isolated disk galaxy into a quasi-steady state where gravitational collapse and cooling are balanced by stellar-feedback-driven turbulence and heating.
In the second stage, we turn off the feedback and zoom into the ISM in various locations where we inject energy and mass to model the evolution of SNRs in a self-consistently generated galactic ISM.

\subsection{Setup: Isolated disk galaxy}
\begin{figure}
 \includegraphics[width=\linewidth, clip=true]{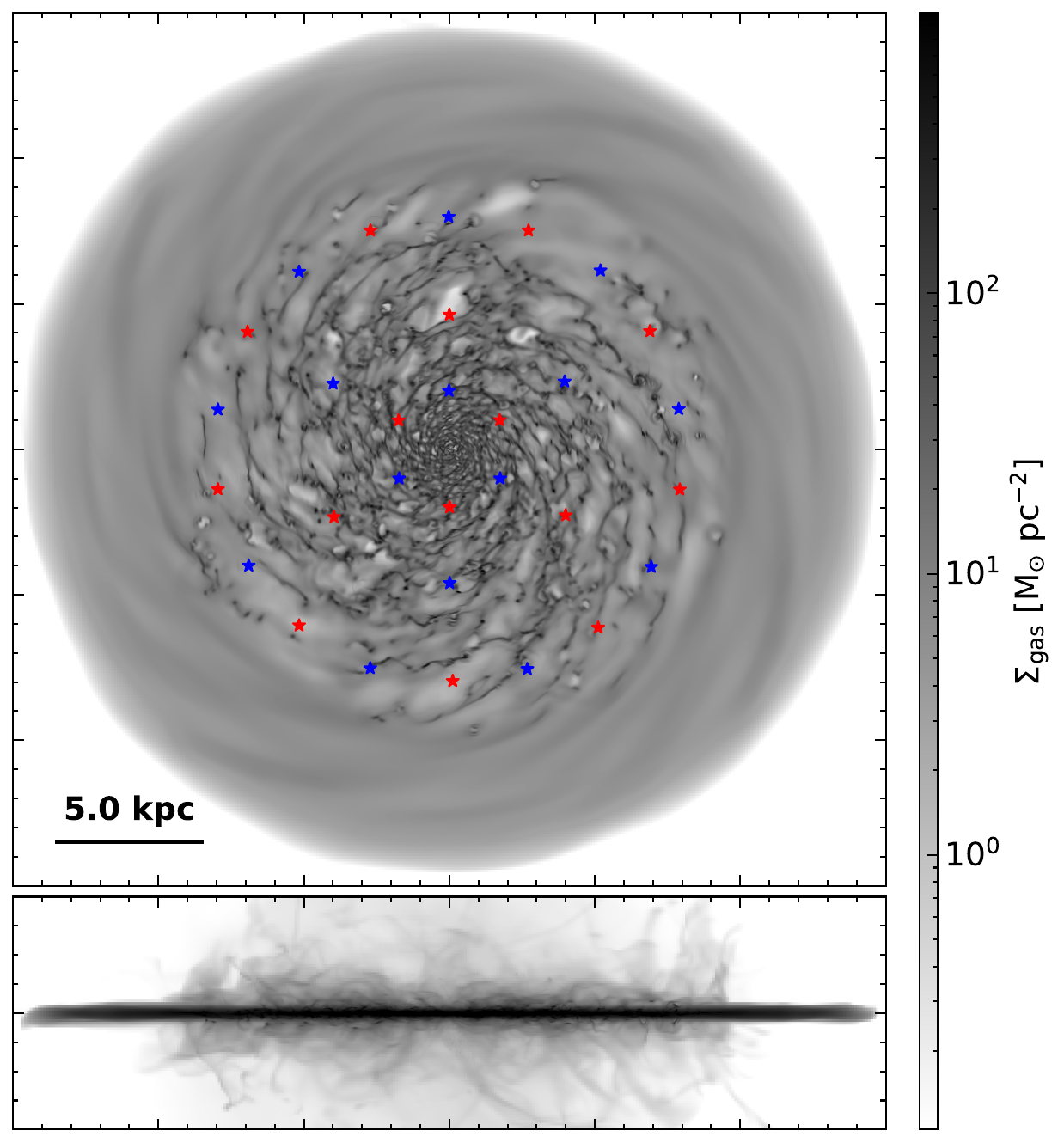}
 \caption{Face-on (top) and edge-on (bottom) projection of the simulated galaxy at $t = 0$. We mark the explosion sites of the SNRs with star markers. Different marker colors correspond to the different passive scalars associated with the SN ejecta. The ISM in the inner $\sim 10\, \text{kpc}$ is highly structured with filamentary outflows that reach several kpc above the midplane, while the ISM in the outskirts is rather smooth without any prominent vertical features.} 
 \label{fig:disk_projection}
\end{figure}

The SISSI galaxy is part of the AVALON galaxy formation and evolution project (M. Behrendt et al. 2025, in prep.), which utilizes the {\sc galaxy composer} package (M. Behrendt et al. 2025, in prep.) to generate the initial conditions of an isolated Milky-Way-like galaxy with galaxy parameters taken from \citet{2016ARA&A..54..529B}. 
The simulation domain is a cubical box with side length $L = 48\, \text{kpc}$ and outflow boundaries, subdivided into a coarse grid of 256 cubic cells, corresponding to a maximum cell size of $\Delta x_{\text{max}} = 187.5 \, \text{pc}$.
Cells are refined up to an effective resolution of $2^{12}$ ($l_{\text{max, ISM}} = 12$) or $\Delta x_{\text{min, ISM}} \approx 11.7 \, \text{pc}$ if they are larger than $N_{\text{Jeans}} = 8$ local Jeans lengths or if they contain a mass exceeding 20 (star) particle masses, which ensures that star-forming cells are Jeans-unstable.
We model the influence of the stellar disk, bulge and dark matter halo as a static, axisymmetric background-potential.
The gas is initially set up as a combination of a warm, isothermal disk in vertical hydrostatic equilibrium and a hot, diffuse uniform background.

During the initial relaxation stage, we model stellar feedback by injecting a thermal energy of $2 \times 10^{52}\, \text{erg}$ and a mass of $200 \, \text{M}_{\odot}$ into a single cell hosting a star particle 8 Myr after its formation.
We avoid overcooling by flagging cells affected by stellar feedback with a passive scalar that disables cooling for the first $\sim 500 \, \text{kyr}$ after the feedback event.

We evolve the isolated disk galaxy for $\lesssim 500 \, \text{Myr}$ until it has settled into a quasi-steady state where gravitational collapse and cooling are balanced by feedback-driven heating and turbulence. 
We show a projection of the surface density of the ISM after the initial relaxation in Fig. \ref{fig:disk_projection}.
Shown here is only a small cut-out of the simulation domain focusing on the galactic ISM. While the large box size is required to ensure a realistic galactic eco-system (galactic outflows and large-scale fountain-flows) and reduce numerical effects due to the domain boundaries, for this study the details of the circumgalactic medium can be ignored, as our focus lies mainly on the central $\sim 8 \,\text{kpc}$.

\subsection{Zoom-in: Treatment of supernova remnants}\label{sec:zoom-in}
\begin{figure}
 \includegraphics[width=\linewidth, clip=true]{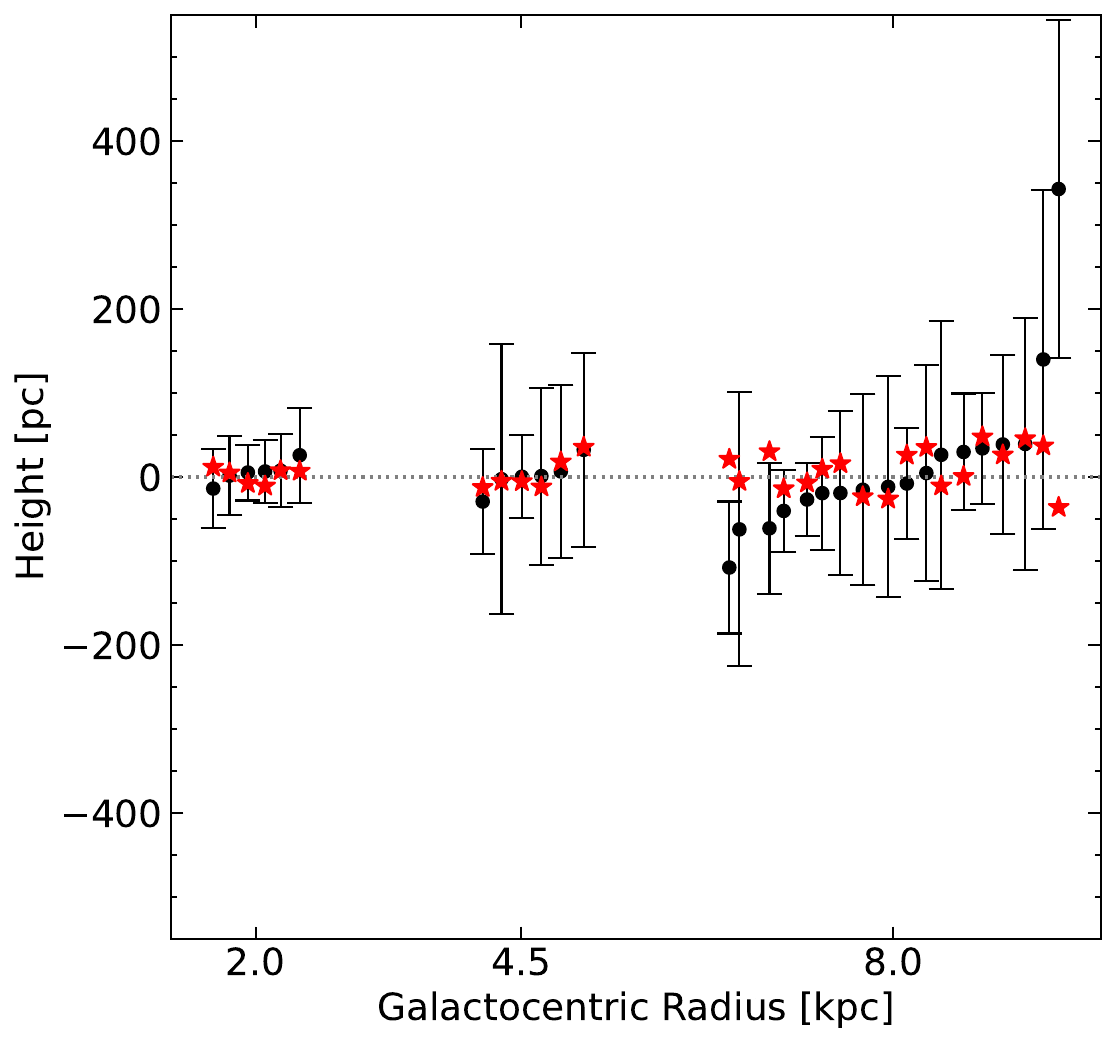}
 \caption{Initial vertical height of the explosion sites, grouped by galactocentric radius (star markers). Black dots denote the local galactic midplane; error bars the vertical scale height defined in the App. \ref{app:ISM}. Radial coordinates, corresponding to $R = 2,\, 4.5\, \text{and}\, 8\,\text{kpc}$, were shifted for visibility. Even though the explosion sites were chosen to be close to $z = 0$, due to the warping of the disk, some of the SNRs are located outside the midplane.} 
 \label{fig:initial_height}
\end{figure}
\begin{figure}
 \includegraphics[width=\linewidth, clip=true]{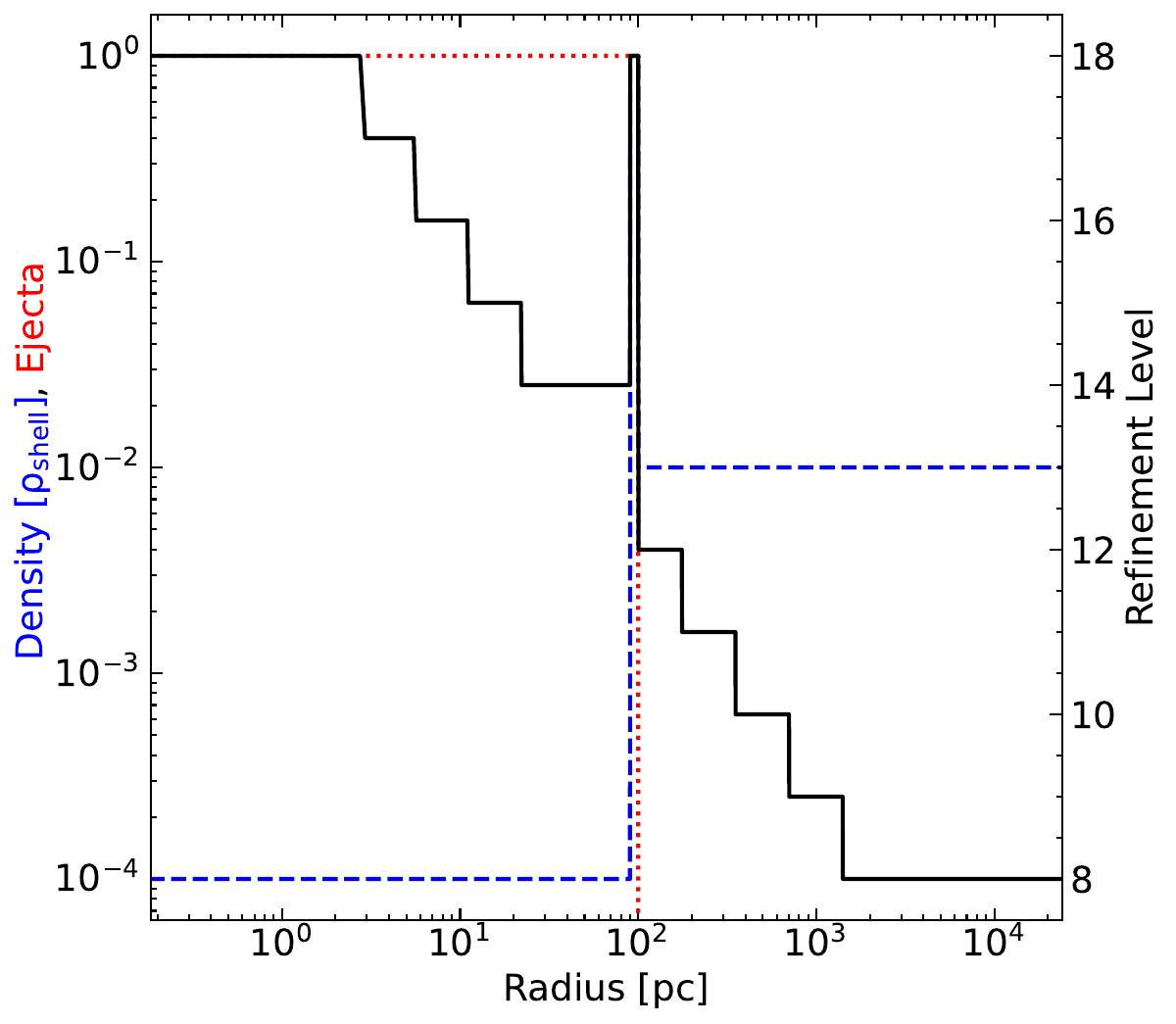}
 \caption{Refinement map produced with the refinement method outlined in Sect. \ref{sec:zoom-in} for the idealized situation of a diffuse bubble with a dense shell, designed to roughly resemble an SNR after shell formation. The solid-black, dashed-blue and dotted-green lines show the radial profiles of the refinement level, gas density and ejecta fraction (scalar tracer field), respectively. The resolution is decreasing radially outward, levels off at $l_{\text{min, zoom}} = 14$ and increases again to $l_{\text{max, zoom}} = 18$ inside the shell.} 
 \label{fig:refinement_ideal}
\end{figure}

We flag 30 star particles at three galactocentric radii $R \in \left\{2, 4.5, 8\right\} \, \text{kpc}$ and $z \sim 0$, spaced equidistantly in polar direction as \textit{SNR} particles (see markers in Figs. \ref{fig:disk_projection} and \ref{fig:initial_height}). An overview of the local ISM properties in the selected regions is given in the App. \ref{app:ISM}.

We refine all cells within $r_{\text{zoom}, l} = N_{\text{zoom}} \Delta x_{l}$ of an SNR particle up to a maximum zoom-in resolution of $l_{\text{max}} = 18$, corresponding to $\Delta x_{\text{min}} \approx 0.18 \, \text{pc}$, where $N_{\text{zoom}} = 15$. 
We further relax the system for $\lesssim 50 \, \text{kyr}$ to avoid numerical artifacts due to the sudden refinement. 
Unless specified otherwise, we measure time from the time of the snapshot at the end of this final relaxation step ($t = 0$).

Starting from $t=0$, each SNR particle injects $N_{\text{SN}}$ SNe per injection.
SN injections may happen every $\Delta t_{\text{SN}}$.
Models differ only by the choice of $N_{\text{SN}}$ and $\Delta t_{\text{SN}}$.

Per SN, each SNR particle distributes $E_{\text{SN}} = 10^{51} \, \text{erg}$ of thermal energy and $M_{\text{ej}} = 5 \, \text{M}_{\odot}$ of ejecta mass evenly within a sphere of radius $R_{\text{inj}} = 5 \Delta x_{\text{min}} \approx 0.92 \, \text{pc}$ centered at the SNR particle's position.
In addition each SNR particle injects one of two passive scalars $Z_{\text{ej, i}}$, corresponding to red and blue markers in Fig. \ref{fig:disk_projection}, used to label the mass fraction of SN ejecta and distinguish between the ejecta of neighboring SNRs.
We note that in \citet{2024ApJ...965..168R} we have demonstrated that the later evolution from the ST phase onward is correctly reproduced, in agreement with the results of \citet{2015MNRAS.451.2757W}, who find that different injection methods reproduce the ST phase equally well, provided short enough time steps are used. 
Other studies focusing on the numerical implementation of SN feedback further corroborate this conclusion \citep{2012MNRAS.426..140D, 2015ApJ...802...99K, 2018MNRAS.477.1578H}.

\begin{figure}
 \includegraphics[width=\linewidth, clip=true]{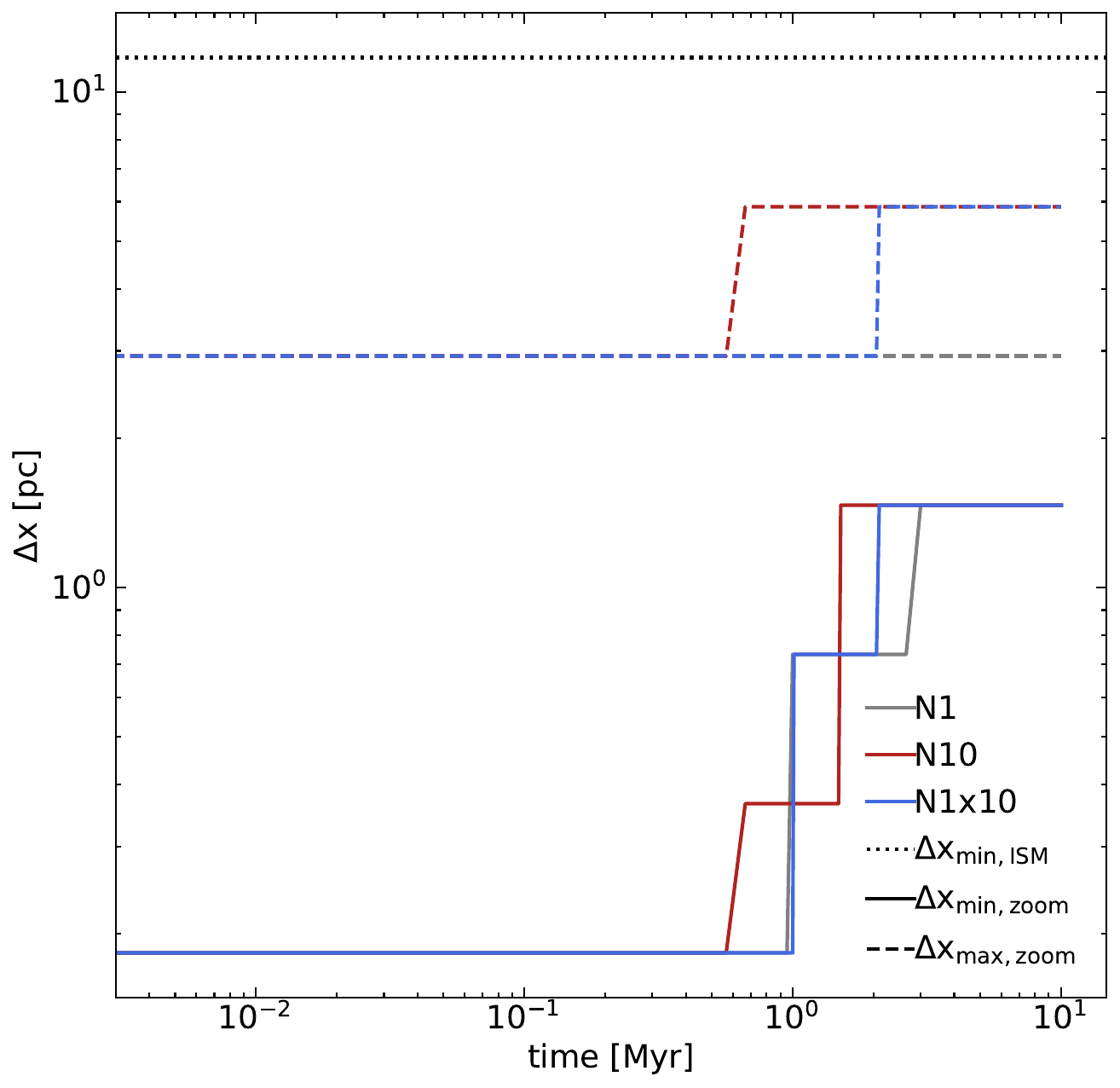}
 \caption{The resolution in the zoom-in region as a function of time. Gray, red and blue lines correspond to the three different runs, while different linestyles correspond to different refinement parameters. 
 The resolution was decreased between restarts of the simulation when the memory requirements became too large. The maximum resolution in the refinement regions around the central SNR particles was left untouched.} 
 \label{fig:resolution_time}
\end{figure}

We refine polluted cells with $Z_{\text{ej, i}} > 10^{-15}$ to at least $l_{\text{min, zoom}}$ and even further up to at most $l_{\text{max, zoom}}$ if
\begin{equation}
    \Delta x > 0.1 \, R_{\text{sf}}^{\text{KO15}}\left(n_{\text{H}}\right) = 2.3 \, \left(\frac{n_{\text{H}}}{\text{cm}^{-3}}\right)^{-0.42} \, \text{pc},
\end{equation}
which roughly resembles the convergence criterion proposed by \citet{2015ApJ...802...99K}.
We show an idealized refinement map in Fig. \ref{fig:refinement_ideal}.
We initially set the zoom-in resolutions to $l_{\text{min, zoom}}=14$ and $l_{\text{max, zoom}} = 18$, corresponding to $\Delta x_{\text{max, zoom}} \approx 2.9 \, \text{pc}$ and $\Delta x_{\text{min, zoom}} \approx 0.18 \, \text{pc}$, respectively, and reduce the resolution as the SNRs grow in order keep the numerical cost at a manageable level as shown in Fig. \ref{fig:resolution_time}.
With this refinement prescription, we are thus able to resolve the momentum generation during the Sedov-Taylor phase and the transition to the radiative phase for a single SN up to densities of $n_{\text{H, max}} \sim 430 \left(\Delta x / \Delta x_{\text{min, zoom}} \right)^{-2.4}\, \text{cm}^{-3}$ \citep[see e.g.][]{2015ApJ...802...99K, 2024ApJ...965..168R}.

With our implementation of star-formation, the mass of star particles formed at higher resolution needs to be adjusted in order to ensure that stars are forming if and only if cells are Jeans-unstable and fully refined.
This condition is satisfied by scaling $m_{\star, l}\propto \Delta x_{l}$.

\subsection{Simulation suite: An overview}

Our simulation suite consists of \textit{four} different runs: A baseline simulation without SNe (N0) and three simulations with SNe labeled N1, N10 and N1x10, corresponding to ($N_{\text{SN}}$, $\Delta t_{\text{SN}}$) = (1, $\infty$), (10, $\infty$) and (1, 1 Myr), respectively. In particular, the models N1 and N10 feature a single explosion event at $t=0$, while only in N1x10 there are subsequent explosion events every $\Delta t_{\text{N1x10}} = 1\, \text{Myr}$.

In N0 no zoom-in is applied. In order to estimate the effect the refinement might have, we have run a 5th simulation labeled N0\_zoom without SNe, but with $N_{\text{zoom}} = 85$.

\section{Analysis}\label{sec:analysis}
\subsection{Classification of ISM components}\label{sec:classification}

In order to be able to meaningfully analyze the SNRs' properties we need to reliably differentiate between SNRs and the unperturbed ISM. Moreover, we classify different components of the SNRs, similarly to the approach of \citet{2024ApJ...965..168R} for a single SNR in a uniform ISM.

We adopt the same method of using the passive scalars to flag cells belonging to an SNR.
Neighboring SNRs inject different passive scalars, which enables us to resolve ambiguities if the SNRs approach or even overlap.
Na\"ively, each SNR corresponds to the set of cells polluted with the respective scalar that are closest to its center (e.g. the corresponding SNR particle).
However, in practice since some SNRs get significantly larger than others, we find that this simple prescription would lead to a large number of cells being grouped incorrectly once the SNRs become too large.
We avoid this problem by creating a weighted Voronoi-tesselation in face-on projection with cells centered at the position of the SNR particles and assigning weights, such that all polluted cells belonging to an SNR lie within the corresponding cell.
We assign these weights by visual inspection.

As opposed to the case studied in \citet{2024ApJ...965..168R}, here, the ISM into which the SNRs are expanding is undergoing constant change. 
Thus, in order to study how the properties of the SNRs depend on the that of the ISM, we need to find an appropriate definition of the \textit{local ISM}.
Here, we define the local ISM as the contents of the smallest rectangular box, containing the entire SNR at all times.
The \textit{unperturbed, local ISM}, then corresponds to the contents of the local ISM without the SNRs, which necessarily shrinks as the SNRs grow. 
This leads to the slight bias, that once a region is swept up by the SNR, it ceases to contribute to the description of the unperturbed ISM. 
Nonetheless, the instantaneous state of the immediate surroundings of the SNRs describes the unperturbed ISM much more accurately than its state at a single point in time (i.e. at $t=0$ or at the time of observation).
We note that while in this study, we do not make use of this definition, future analyses using the SISSI simulations will (L. Romano et al., in prep).

We further classify different components of the unperturbed ISM and the SNRs.

For the SNRs we follow the classification of \citet{2024ApJ...965..168R}.
We distinguish between radially inflowing and outflowing shell and bubble components.
The \textit{bubble} corresponds to polluted, hot ($T > 2 \times 10^{4}\,\text{K}$) or diffuse ($n_{\text{H}} < 10^{-2}\, \text{cm}^{-3}$) gas, while the \textit{shell} corresponds to cold and dense, polluted gas.
We decide whether the gas is in- or outflowing by measuring the radial velocity, measured from the center of mass of the SNR in the co-rotating, center-of-mass frame of each SNR.

For the unperturbed ISM we distinguish between \textit{cold} ($T < 7 \times 10^{3}\, \text{K}$), \textit{warm} ($7 \times 10^{3}\, \text{K} < T < 10^{5} \, \text{K}$) and \textit{hot} ($10^{5} \, \text{K} < T$) gas phases, which are expected to coexist co-spatially in a turbulent medium with inhomogeneities driven by SN explosions and differential cooling \citep[see e.g.][]{1977ApJ...218..148M, 2005ARA&A..43..337C}. The choice of $7 \times 10^{3}\, \text{K}$ for the threshold between warm and cold gas, slightly less than the commonly used $\sim 10^4\, \text{K}$, arises from the adopted cooling function, which produces persistent gas at this temperature as is shown in the App. \ref{app:ISM}.

We also classify the stars within the ISM boxes based on whether they are \textit{old}, i.e. formed before $t=0$ or \textit{young}.
For the \textit{young} stars we further distinguish between stars that are formed from \textit{polluted} or \textit{pristine} gas.

\subsection{Definition of polluted cells}\label{sec:pollution}

We define a cell to be polluted if its passive scalar concentration exceeds some threshold value $Z_{\text{ej, thr}}$.
The choice of this threshold value is arbitrary and can systematically bias our results.
If we choose a value of $Z_{\text{ej, thr}}$ that is too low, we risk including gas that is only (slightly) polluted due to numerical noise, but that physically is not associated with the SNRs.
On the other hand if we choose a value that is too high, we risk missing parts of the SNRs.

In practice it seems impossible to entirely prevent both effects from happening, so we aim for a compromise and state our results in terms of range of plausible values based on a slightly high and a slightly low threshold value.
We first perform our analysis for a slightly low value $Z_{\text{thr, low}} = 10^{-12}$, comparable to the value used in \citet{2024ApJ...965..168R}.
After defining the local ISM boxes, based on the SNRs defined by the choice of $Z_{\text{thr, low}}$, we define $Z_{\text{thr, high}}^{i}\left(t\right)$ for each SNR and snapshot, by requiring that the total ejecta mass of cells with $Z_{\text{ej, i}} > Z_{\text{thr, high}}^{i}\left(t\right)$ just exceeds 99.99 per cent of the total gas-phase ejecta-mass in the ISM box.

\subsection{SNR geometry}\label{sec:ellipsoid}

We study the dynamical evolution of the SNRs' geometry, by analyzing how their shape tensors evolve over time. We define the shape tensor as
\begin{equation}
    S_{ij} = V_{\text{SNR}}^{-1}\int_{\text{SNR}} \left(\left\lVert \mathbf{x}\right\rVert^2\delta_{ij} - x_{i}x_{j}\right)\text{d}^3\mathbf{x} ~,
\end{equation}
which is the volume weighted inertia tensor, assuming a constant density of unity.
By assuming an approximately ellipsoidal shape, we can define the three ellipsoidal radii, defined as
\begin{equation}
    r_{i} = \sqrt{2.5\left(\text{tr}\left(S\right) - 2S_{i}\right)} ~,
\end{equation}
where $S_{i}$ are the eigenvalues of $S_{ij}$ and $\text{tr}\left(S\right)$ the trace.
We refer to the smallest, intermediate and largest eigenvalue as the minor $a$, semi-major $b$ and major $c$ axis, respectively.
We define the effective size of an SNR as the geometric mean of the three eigenvalues
\begin{equation}
    r_{\text{eff}} = \left(a b c\right)^{1/3}~.
\end{equation}

To determine the alignment of the SNRs within the galaxy, we measure their pitch angle $\alpha$ and polar direction $\text{cos}\left(\theta\right)$ for both the major and minor axes.
The pitch angle is defined relative to the direction of galactic rotation, with $\alpha = 90^\circ$ and $\alpha = -90^\circ$ corresponding to the galactic center and anti-center, respectively.
The magnitude of the polar direction is 0 (1) for directions parallel (perpendicular) to the galactic plane. 

\section{Time evolution of SNRs}\label{sec:time_evolution}
\subsection{Showcase: Supernovae in relatively uniform medium} \label{sec:n_0}
\begin{figure*}
 \includegraphics[width=\linewidth, clip=true]{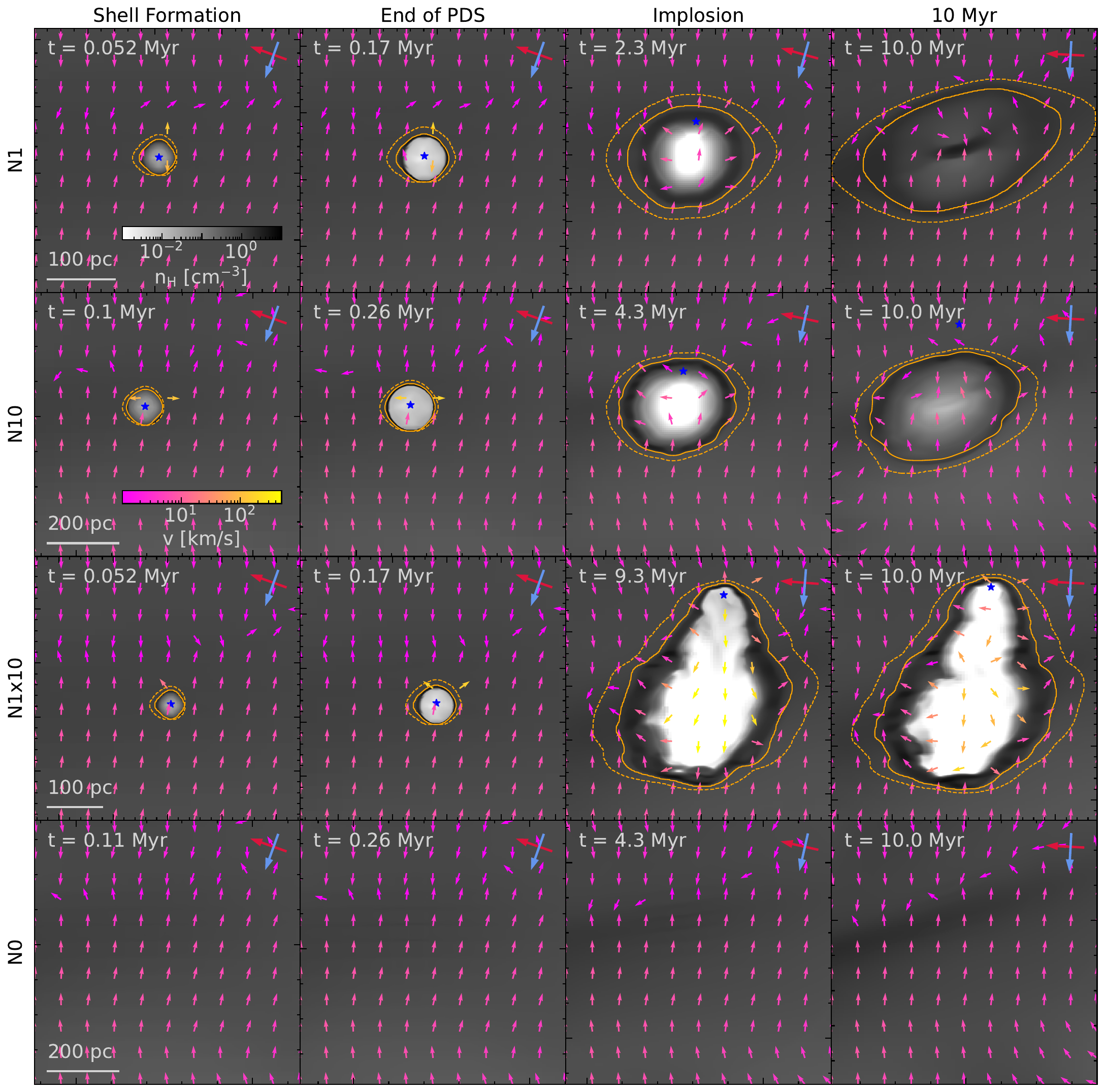}
 \caption{Density-slices through the central plane of the SNR \#22 at various points in time for each model. Arrows are depicting the velocity field in the co-rotating center-of-mass frame of the local ISM.
 The various timescales correspond to different points in time for the different models. 
 Red and blue arrows in the top-right corner of each panel indicate the directions of the galactic rotation and the galactic center, respectively. The dashed orange contour corresponds to the surface where $Z_{\text{ej}} = Z_{\text{thr, low}}$, while the solid contour corresponds to $Z_{\text{ej}} = Z_{\text{thr, high}}$.
 Since the various timescales are undefined for the model no\_expl, we are using the same times as model N10. 
 The SNe explode into a fairly homogeneous ISM, with a slowly collapsing, slight overdensity right where the SNe explode.
 At similar evolutionary stages the SNR is about twice as large in N10 compared to N1, with very similar geometry; Spheroidal with a slight elongation in the direction of rotation.
 On the other hand the geometry in the model N1x10 qualitatively differs from the other models, with an elongated cavity normal to the rotational direction, due to the elliptical orbit ($v_{\text{R}} \sim 20\, \text{km/s}$) of the explosion site.
 Only in the model N1, after 10 Myr a dense cloud, aligned with the SNR is forming in the center as predicted by \citet{2024ApJ...965..168R}.} 
 \label{fig:slices}
\end{figure*}
\begin{figure*}
 \includegraphics[width=\linewidth, clip=true]{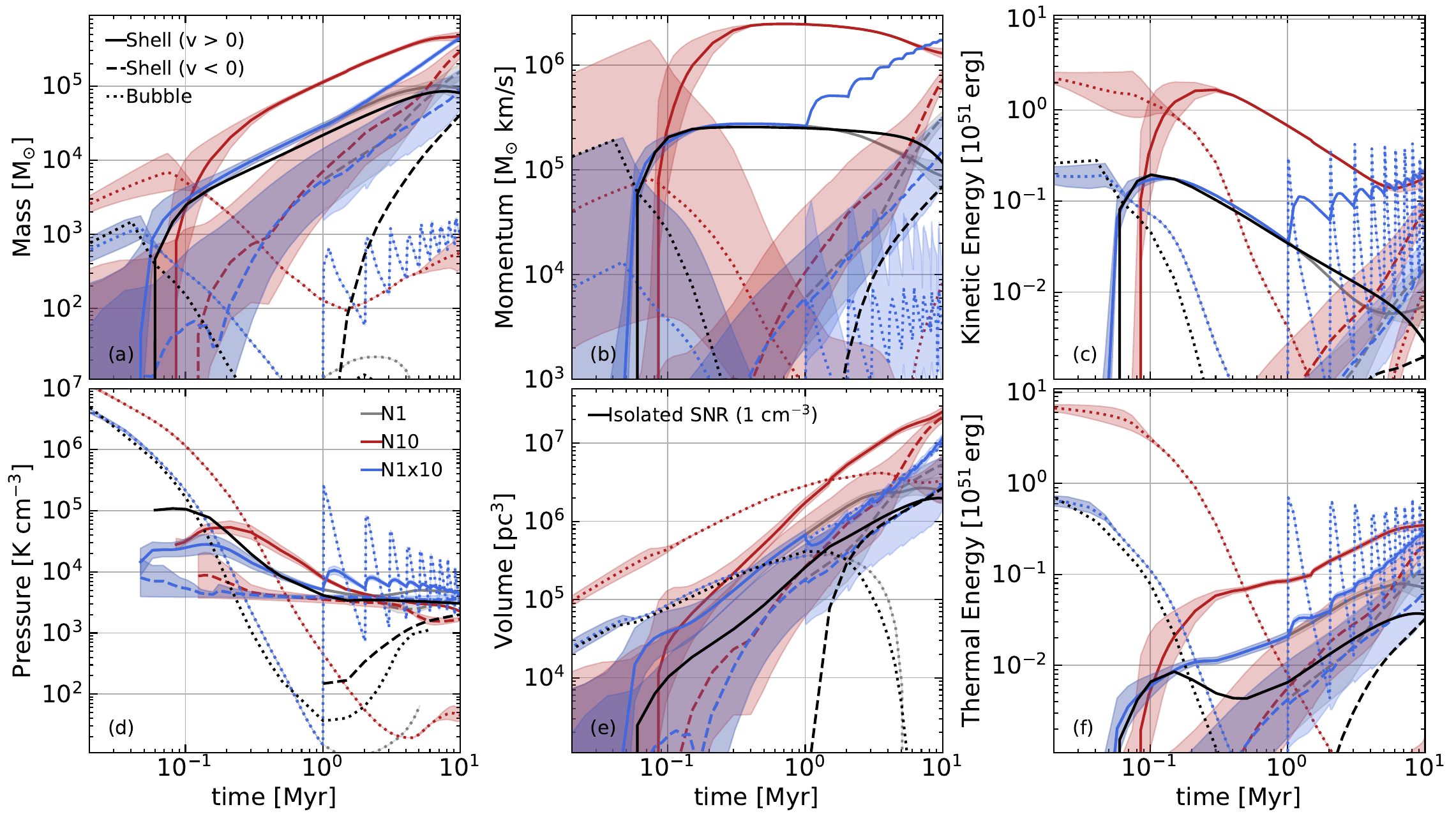}
 \caption{Various properties (Mass, momentum, kinetic and thermal energy, pressure and volume) of the SNR \#22 as a function of time for the various models using the classification introduced in Sec. \ref{sec:classification}. Shaded regions correspond to the margin of uncertainty introduced by the choice of $Z_{\text{ej, thr}}$, while the lines correspond to the geometric average of the values obtained with high and low values of $Z_{\text{ej, thr}}$.
 For comparison we show the time evolution of an isolated SNR at a similar ambient density ($n_{\text{H}} = 1 \, \text{cm}^{-3}$) taken from \citet{2024ApJ...965..168R}.
 As expected, the models N1 and N10 exhibit similar behavior and the model N1 also agrees quantitatively quite well with the isolated SNR.
 In the model N1x10, the SNR initially follows the model N1 and then after the onset of the consecutive SNe diverges reaching a comparable mass, momentum and size as the model N10 after 10 Myr.
 However, the fraction of thermal energy in the bubble is higher in N1x10 compared to N10, indicating more efficient hot phase generation.
 Difference due to the choice of $Z_{\text{ej, thr}}$ are largest before shell formation and are most pronounced in the mass and momentum of the bubble, indicating that ejecta are initially lagging behind the shock, but catch up once a cold shell forms.} 
 \label{fig:global_evolution}
\end{figure*}

The case of stellar feedback in an ambient medium with solar metallicity and an ambient density of $n_{\text{H}} \sim 1\,\text{cm}^{-3}$ has been widely studied \citep[e.g.][]{2015ApJ...802...99K, 2016MNRAS.456..710F, 2022ApJS..262....9O, 2024ApJ...965..168R}.
In this section we showcase the results of SNR \#22, which happens to explode in a relatively uniform medium with an ambient density close to $1\,\text{cm}^{-3}$ and compare its time evolution to that found in previous studies.

In Fig. \ref{fig:slices} we show slices of the density field through the center of SNR \#22 parallel to the xy-plane at various characteristic times for the different models.
In each panel, the outline of the SNR is shown by orange lines, depicting contours of constant $Z_{\text{ej}}$, corresponding to $Z_{\text{thr, low}}$ (dashed line) and $Z_{\text{thr, high}}$ (solid line).

As can be seen in the bottom row, corresponding to the N0 model, the density field is indeed rather uniform, but some collapse into a filamentary structure over several Myr is visible.

At shell formation (first column) the SNRs are spherical with a slightly underdense central region and a thin, overdense shell. The time of shell formation and the SNRs' sizes are in agreement with previous work \citep[e.g.][]{2015ApJ...802...99K}.

After shell formation, SNRs enter the so-called pressure-driven snowplow (PDS) phase, which ends once the pressure in the cavity drops below that of the shell (second column).
At this time, the SNRs are spherical, with an increasingly underdense central region and a thin, overdense shell. 
The time at which the PDS phase ends and the SNRs' sizes are in agreement with previous estimates \citep{2024ApJ...965..168R}.

\citet{2024ApJ...965..168R} have shown that SNRs implode as they merge with their ambient medium.
In their simulations, a SNR in an ambient medium with $n_{\text{H}} \sim 1\,\text{cm}^{-3}$, such as the one considered here, began to implode after $\sim 1 \, \text{Myr}$.
In the third column we show the SNRs right after the onset of implosion.
By this time the SNRs are slightly elongated, parallel to the collapsing filament, which is at a slight angle to the direction of galactic rotation.
In all cases, the implosion occurs significantly later than our expectation based on previous work.
In the model N1x10, the implosion seems to be coincident with the explosion happening at $t \sim 9\,\text{Myr}$ and is no longer visible by $t = 10\, \text{Myr}$. We rule out this ``implosion'' as a false positive and caution that with our definition of the implosion timescale we cannot distinguish between brief moments of radially ``inflowing'' ejecta in N1x10 due to a displacement of the explosion sites and sustained inflows of cold gas from the shell.

After 10 Myr (fourth column), the SNRs in models N1 and N10 have been stretched out considerably in the direction of the collapsing filament. The implosion in N1 has reached the center and condensed into a growing, filamentary implosion cloud, as predicted by \citet{2024ApJ...965..168R, 2024ApJ...971L..44R}, which is stabilized by the rapid radiative dissipation of the energy carried by the implosion shocks colliding in a central region. 
In N10 the center of the SNR is still underdense indicating that the implosion has not yet reached the center.
Meanwhile, in N1x10 the SNR is stretched out predominantly in radial direction following the wake of the explosion center, which happens to be drifting radially outward. The interior of the superbubble remains strongly underdense.

One can see, that the volume traced by the dashed line corresponding to $Z_{\text{thr, low}}$ tends to be slightly larger than the SNRs, particularly along the directions aligned with the Cartesian grid at early times, and at late times the direction of galactic rotation.

In Fig. \ref{fig:global_evolution} we show various global properties of the SNRs as a function of time and compare them to those of a single SN exploding into a uniform medium with an ambient density of $n_{\text{H}} = 1\,\text{cm}^{-3}$ taken from \citet{2024ApJ...965..168R}.
We note that they used a different cooling function, leading to a slightly lower equilibrium pressure (see panel (d)).

All quantities are defined in essentially the same way as in \citet{2024ApJ...965..168R}, i.e. extensive quantities are computed by summing up the contributions from all cells belonging to the respective gas phases and the pressure is calculated as a volume weighted average.
However, due to the differential movement of the ambient medium, differences might arise in the momentum and kinetic energy due to the choice of inertial system, which we here have chosen to be the center-of-mass system after subtracting the galactic rotation, azimuthally averaged in linearly spaced, radial bins with spacing $\Delta R \approx 47 \, \text{pc}$.

In the model N1, all quantities except the pressure and the thermal energy agree with the isolated SNR for the first $\sim 2 \text{Myr}$ within uncertainties.
At late times the radial momentum and kinetic energy drop more rapidly. 
The kinetic energy, eventually recovers and levels off at $\sim 1\, \%$ of the injected explosion energy.
However, these small differences might well be explained by the choice of the inertial system.

Model N10 appears to be a rescaled version of N1, in line with the idea, that SNRs undergo a series of self-similar evolutionary stages.

After 1 Myr, model N1x10 starts to diverge from N1.
The amount of swept up mass, the total radial momentum, and kinetic as well as thermal energy of the SB grow to be quite similar to those of N10, at $t = 10\,\text{Myr}$; though with large temporal variations in the distribution between the bubble and the shell.
This indicates that these quantities are mostly sensitive to the total amount of injected energy, regardless of the exact interval between injections, in stark contrast to the geometry and mass distribution within the SNR, as shown in Fig. \ref{fig:slices}. Nonetheless, in model N1x10 the SNR is retaining more radial momentum, and sustaining a higher pressure, indicating that the subsequent radiative losses are determined by the multi-phase structure of the ISM at the injection site.

\subsection{The full SISSI sample} \label{sec:full_sample}

\begin{figure*}
 \includegraphics[width=\linewidth, clip=true]{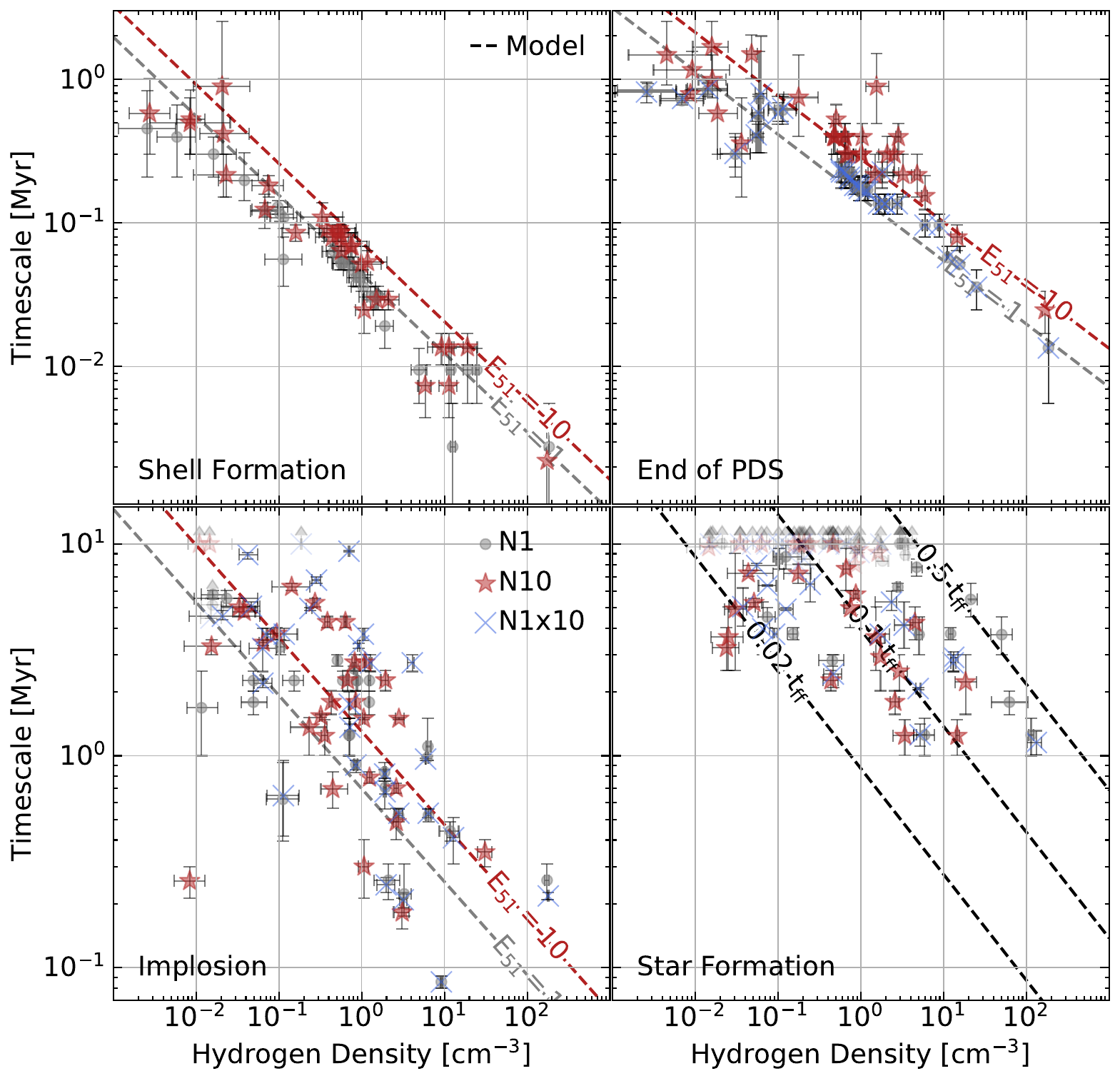}
 \caption{Various timescales as a function of ambient density for our simulated sample of SNRs. Uncertainties arise due to the finite spacing of the snapshots and due to the choice of $Z_{\text{ej, thr}}$. The timescales of shell formation, the end of the PDS phase and SNR implosion agree well with the predictions from models based on simulations of isolated SNRs \citep{2015ApJ...802...99K, 2024ApJ...965..168R}. The timescale measuring the onset of star formation within the SNRs is within a factor of five of 10 per cent of the free-fall timescale of the star-forming SNRs.} 
 \label{fig:timescales}
\end{figure*}

\begin{figure*}
 \includegraphics[width=\linewidth, clip=true]{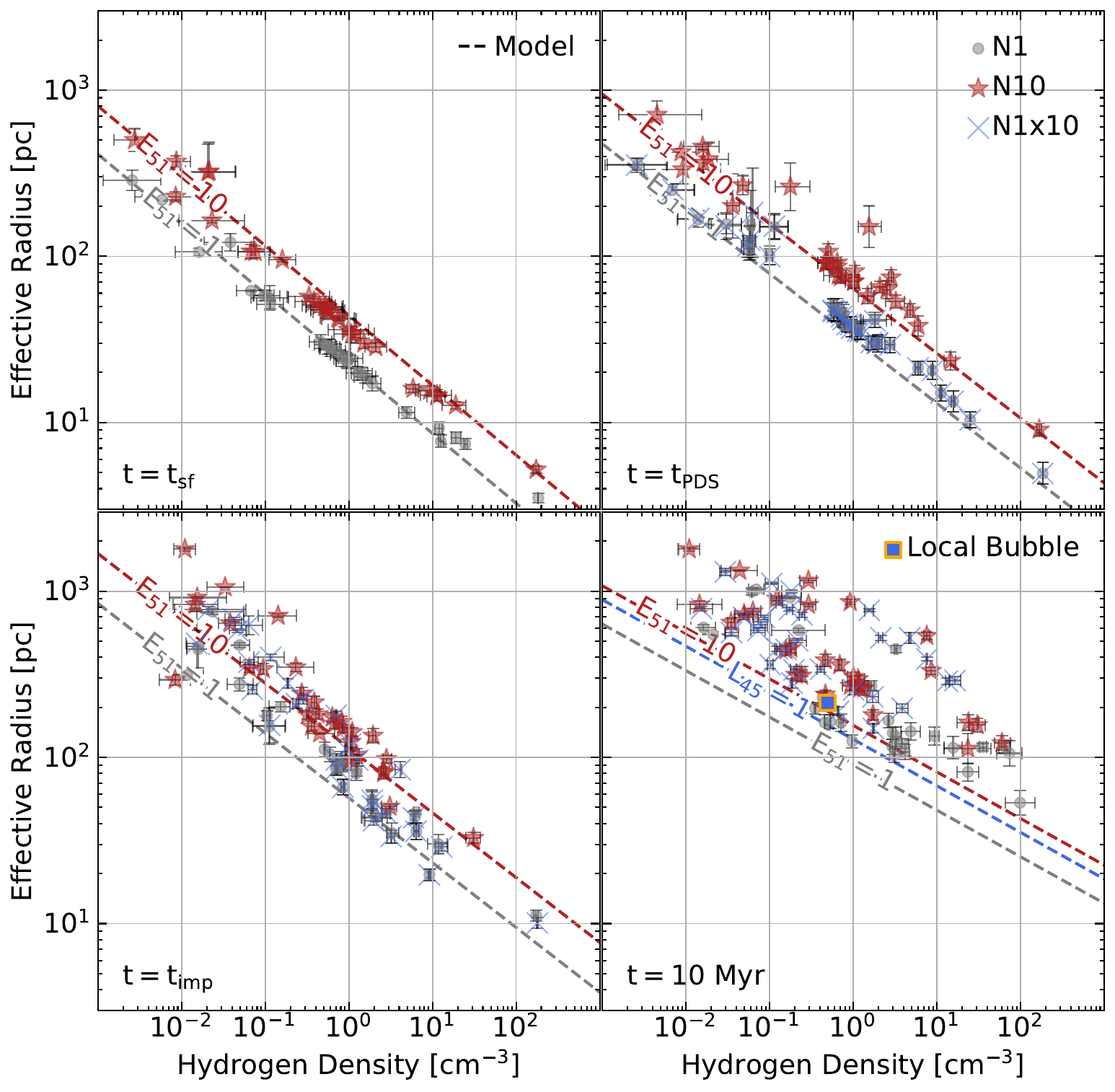}
 \caption{SNR size at various characteristic points in time as a function of ambient density for our simulated sample of SNRs. Uncertainties arise due to the finite spacing of the snapshots and due to the choice of $Z_{\text{ej, thr}}$. 
 An orange and blue square depicts the effective radius of the LB derived from the 3D dust maps of \citet{2024A&A...685A..82E} in the panel corresponding to $t = 10 \, \text{Myr}$. Error bars are smaller than the marker and thus not shown.
 The radii at shell formation, the end of the PDS phase and at SNR implosion agree well with the predictions from models based on simulations of isolated SNRs \citep{2015ApJ...802...99K, 2024ApJ...965..168R} for sufficiently large ambient densities. At low densities the sizes tend to exceed the model predictions. 
 After 10 Myr the SNRs are about twice the expected size.} 
 \label{fig:sizes}
\end{figure*}

\begin{figure}
 \includegraphics[width=\linewidth, clip=true]{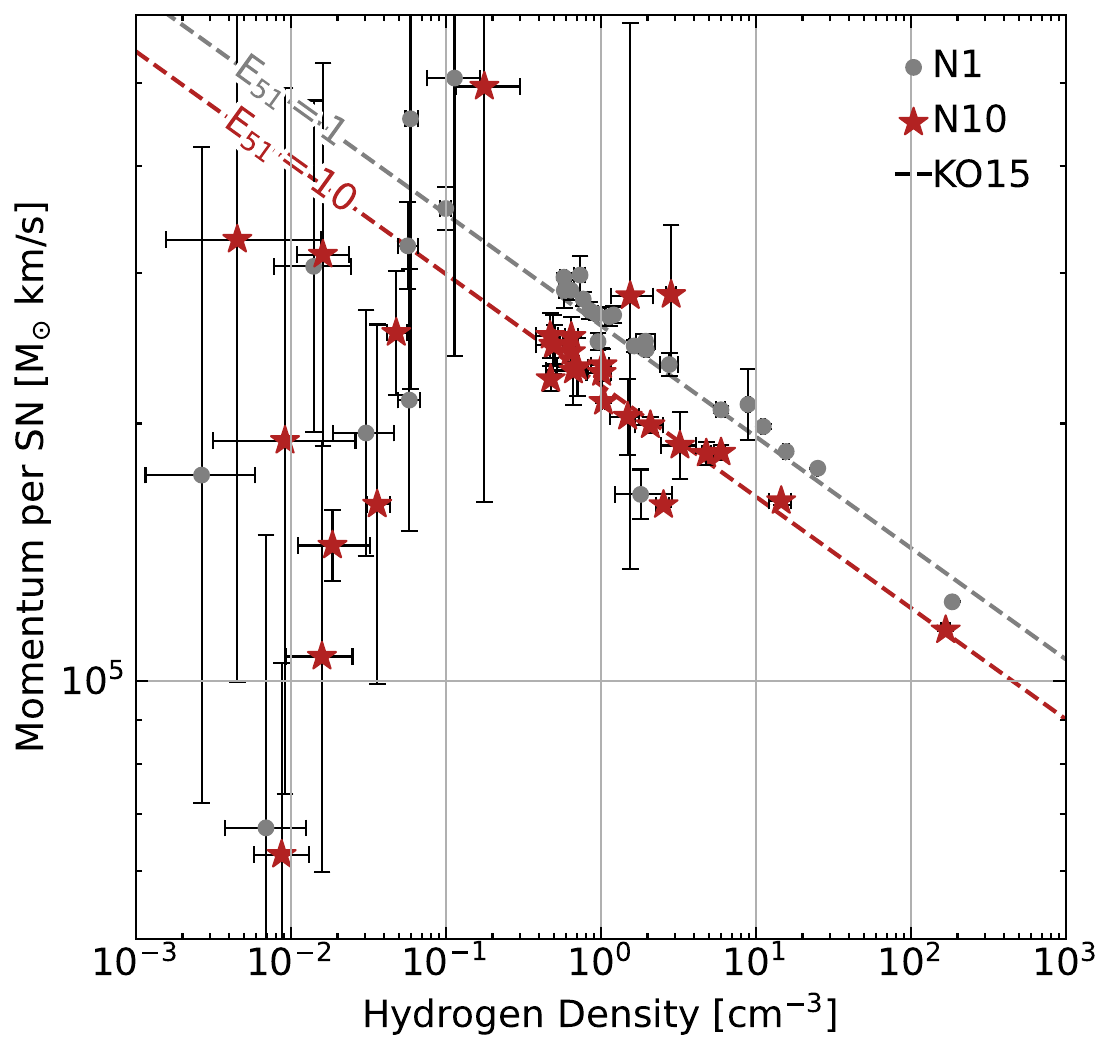}
 \caption{Outward, radial momentum per SN at the end of the PDS phase as a function of ambient density. Uncertainties arise due to the finite spacing of the snapshots and due to the choice of $Z_{\text{ej, thr}}$. At sufficiently high density $n_{\text{H}} \gtrsim 0.1 \, \text{cm}^{-3}$ the momentum input matches well the the prediction of simulations of isolated SNRs \citep{2015ApJ...802...99K, 2024ApJ...965..168R}. On the other hand, at lower densities the momentum per SN is often lower than expected, with large error bars.} 
 \label{fig:momentum}
\end{figure}

We have shown, that SNR \#22 adheres well to the expectations from isolated SNRs in uniform ambient media.
However, this might just have been a special case that cannot be applied to the whole sample. 
Thus, in this section we evaluate to what extent our full sample of SNRs follows the expectations from previous work.

In Fig. \ref{fig:timescales} we show various characteristic timescales as a function of ambient density, defined here as the ratio of the swept-up mass and the volume covered by each SNR at each point in time, for the different models and compare them with analytical results from previous work, described in the App. \ref{app:theory}, shown as red and gray lines in the different panels.

We find that the shell-formation timescale of the simulated SNRs matches the theoretical estimate Eq. \ref{eq:t_sf}, in line with the expectation that SNRs are hardly affected by the galactic environment during the early adiabatic expansion phase.
The same holds true for the timescale for the end of the PDS (Eq. \ref{eq:t_PDS}), with some exceptions at very low densities.

Differences to the purely analytic picture become more apparent when comparing the timescale of implosion Eq. \ref{eq:t_launch}, where we assume $\sigma_{1} = 0.8$ corresponding to an ambient pressure of $P_{\text{ISM}} \sim 10^4\, k_{\text{B}}\,\text{K cm}^{-3}$, matching the pressure of the isobaric phase of the ISM (Fig. \ref{fig:phase_diagram}).
Here we define $t_{\text{launch}}$ slightly differently from \citet{2024ApJ...965..168R}, who defined $t_{\text{launch}}$ as the time of the first snapshot when at least $0.1 \, \text{M}_{\odot}$ are in the form of backflowing shell gas.
In SISSI, this condition would be met at almost all times, due to the uncertainties in the selection of the SNR gas and the turbulent motion of the background medium. 
We thus restrict the criterion to the ejecta, and define $t_{\text{launch}}$ as the earliest time when the backflowing part of the shell contains at least $2\,\%$ of the ejecta, tagged by the respective scalar tracer. 
We find, that while the bulk of SNRs is not too far from the analytical model, there is considerable scatter and a number of extreme outliers. 
Moreover, as we note in the discussion of Fig. \ref{fig:slices}, the interpretation of SNR implosion in the context of model N1x10 is somewhat unclear, as the interior pressure of the SBs tends to remain high.

We also show the time after which stars start to form from material polluted by SNe.
There is no star-formation from polluted gas for the first 1 Myr, but within about a factor of 5 of $0.1\,t_{\text{ff}}$ stars begin to form within the SNRs, where 
\begin{equation}\label{eq:t_ff}
    t_{\text{ff}} = \sqrt{ \frac{3 \pi}{32 G\rho}} \sim 44.9 \, n_{0}^{-0.5} \, \text{Myr}\,,
\end{equation}
is the free-fall timescale.
Importantly, in many cases this star formation does not appear to be triggered within the SNRs, but rather is the continued star-formation in pre-existing star-forming regions, that are swept up and enriched by the SNRs.
A more detailed analysis of the potential triggering of star-formation in SISSI is out of the scope of this work and will be the focus of future publications (L. Romano et al. 2025, in prep.).

In Fig. \ref{fig:sizes} we show the effective size, as defined in Sect. \ref{sec:ellipsoid}, as a function of ambient density for the different models at various characteristic points in time.
Red, gray and blue lines depict the expected sizes, building on the theoretical models described in the App. \ref{app:theory}.
In the panel corresponding to the last snapshot at 10 Myr, the orange and blue square corresponds to the effective size of the LB derived from the data products of \citet{2024A&A...685A..82E} as described in 
a companion paper (L. Romano et al. 2025).

We find, that overall SNR sizes are in line with theoretical expectations during the stages of SNR evolution before merging with the ISM, i.e. before $t=t_{\text{launch}}$, but start growing larger than expected at later times.
SNRs in low density environments $n_{\text{H}} \lesssim 0.1 \, \text{cm}^{-3}$ start to diverge from the theoretical expectation by $t_{\text{imp}} \gtrsim 1\,\text{Myr}$. 

After 10 Myr all SNRs are about twice the expected size, indicating the need for better models of old SNRs in a shearing, stratified ISM.
Interestingly, the LB is on the smaller end of the sizes for simulated SNRs in similar density media, even though it is expected to be older, i.e. $t_{\text{LB}} \sim 14\,\text{Myr}$ \citep{2022Natur.601..334Z, 2006A&A...452L...1B}. 
We further discuss this point and its implications in a companion paper (L. Romano et al. 2025).

In Fig. \ref{fig:momentum} we show the momentum input per SN at the end of the PDS stage and compare it to the theoretically expected value, assuming a momentum enhancement after shell-formation of $\sim 20\,\%$, slightly lower than the $\sim 50\,\%$ reported by \citet{2015ApJ...802...99K}.
The momentum input in the denser regions $n_{\text{H}} \gtrsim 0.1\,\text{cm}^{-3}$ roughly follows the theoretical expectation, with little scatter.
In contrast to lower density regions, where the momentum per SN drops off with large scatter. This behavior is likely due to the large size ($\gtrsim 100\,\text{pc}$) of these SNRs, leading to more frequent energy dissipation due to interactions with high density structures.

\section{Geometry of simulated SNRs}\label{sec:geometry}

\begin{figure}
 \includegraphics[width=\linewidth, clip=true]{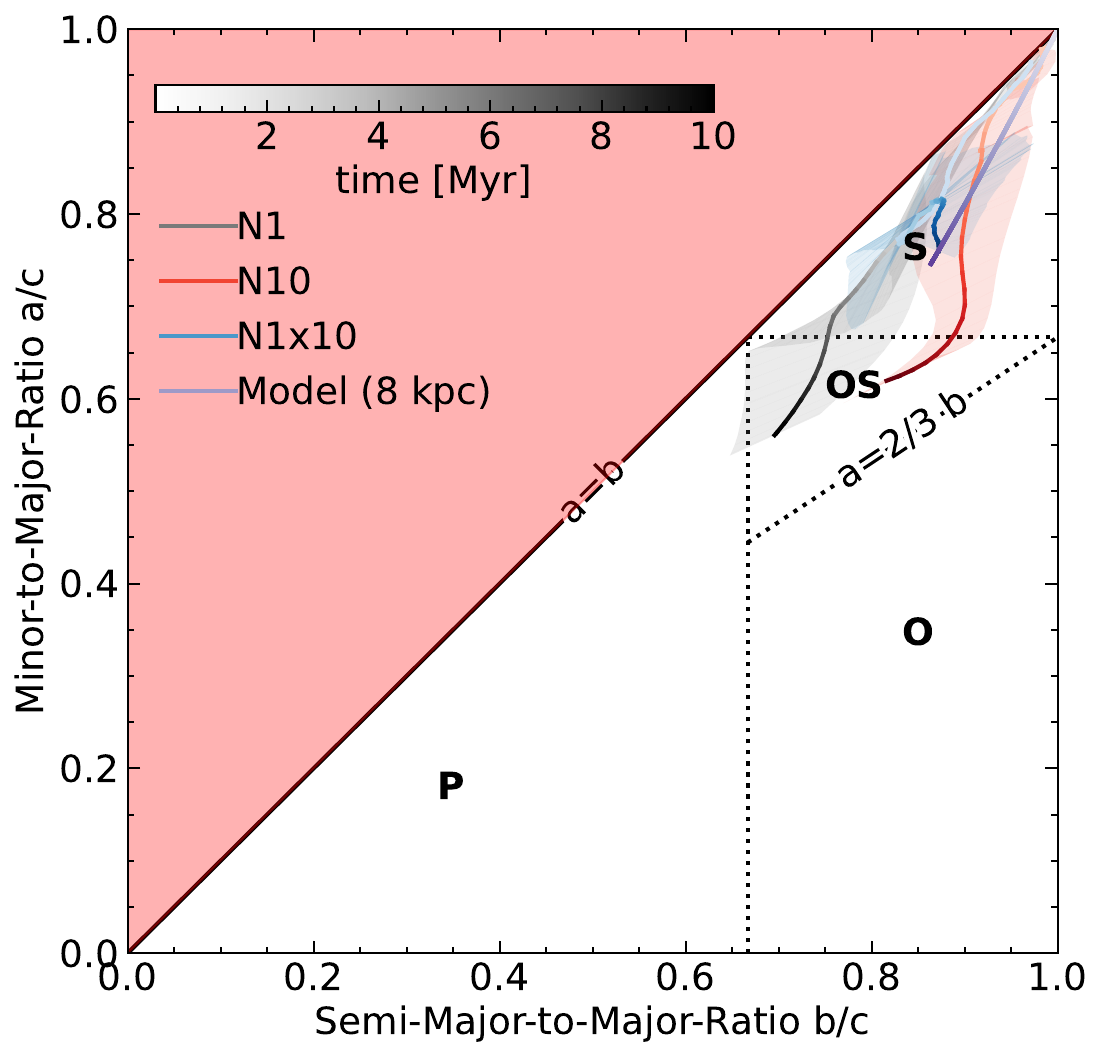}
 \caption{Evolutionary tracks of the SNR \#22 in the shape phase-space for the various explosion models. 
 Uncertainties due to the choice of $Z_{\text{ej, thr}}$ are shown as shaded regions. In different parts of the phase space the SNRs are either spherical (S), oblate spheroids (OS), prolate (P) or oblate (O).
 The SNR starts out as a perfect sphere and becomes increasingly prolate over time. 
 The ratio of the two minor axes remains close to one and never falls below 2/3.
 In the model N10 the SNR remains spherical for longer compared to N1. 
 Similarly the consecutive SN explosions in the model N1x10 restore spherical symmetry.} 
 \label{fig:geometry_track}
\end{figure}

\begin{figure*}
 \includegraphics[width=\linewidth, clip=true]{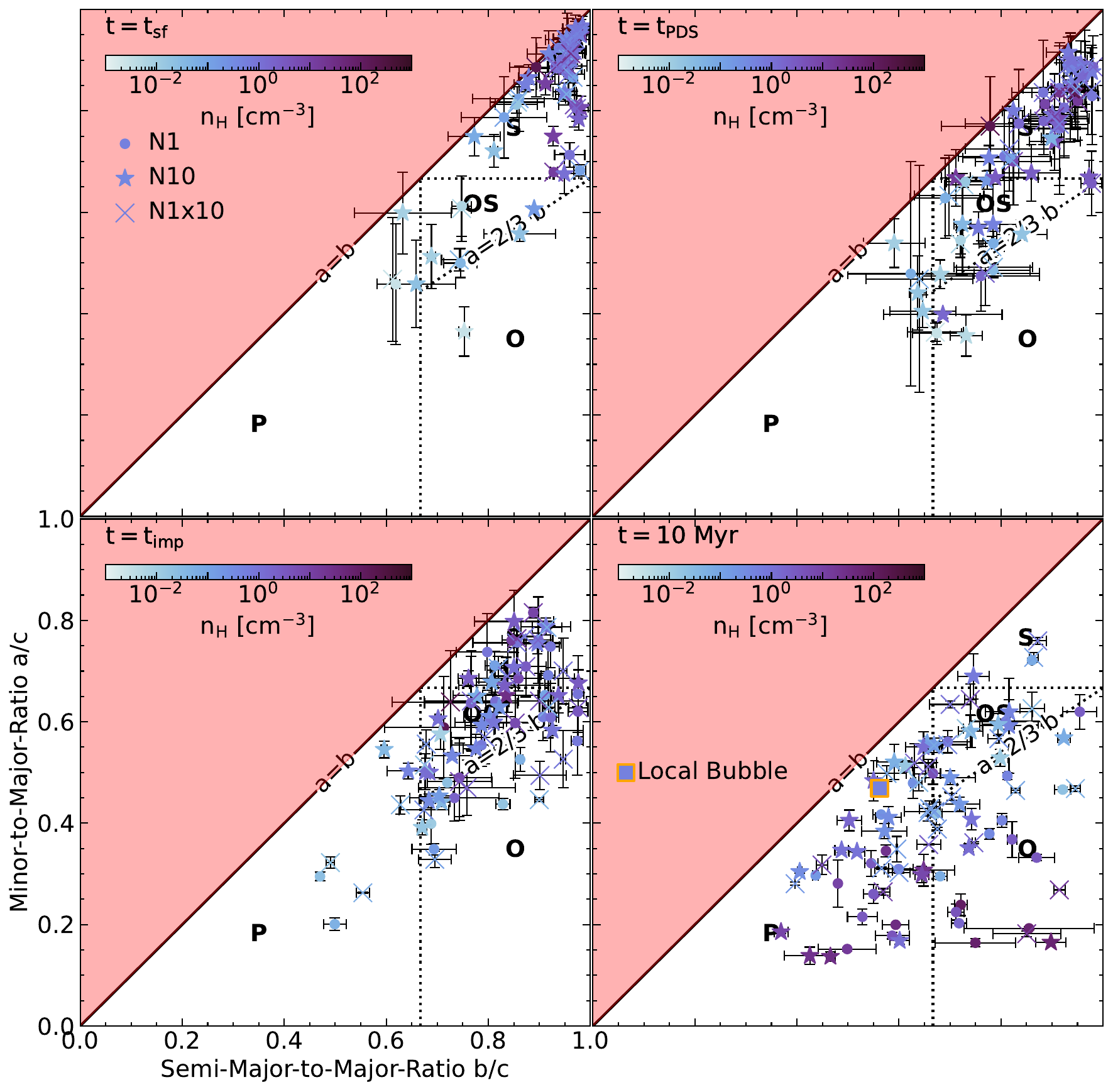}
 \caption{Distribution of SNR shapes at various characteristic points in time for the different explosion models.
 Uncertainties due to the choice of $Z_{\text{ej, thr}}$ are represented by error bars.
 Different regions are labeled as in Fig. \ref{fig:geometry_track}.
An orange and blue square depicts the shape of the LB derived from the 3D dust maps of \citet{2024A&A...685A..82E} in the panel corresponding to $t = 10 \, \text{Myr}$. Error bars are smaller than the marker and thus not shown.
 At shell formation the SNRs tend to be spherical, with SNRs in lower-density environments being somewhat less spherical. At the end of the PDS stage and the onset of the implosion, most SNRs are still spherical or oblate spheroids with $a/b \gtrsim 0.5 - 0.67$. However, some of the lower-density SNRs are already quite asymmetric falling into the prolate and oblate category. The SNRs in the N10 model tend to be somewhat more spherical.
At 10 Myr, the majority of SNRs are asymmetric. In dense environments SNRs tend to be quite asymmetric with low $a/c \sim 0.2$. The model N1 tends to have lower $a/b \sim 0.5$ compared to the other explosion models which tend to have $a/b \gtrsim 2/3$.} 
 \label{fig:snr_shapes}
\end{figure*}

In the previous section we have shown that while young SNRs are well described by the theory based on models in a uniform, stationary medium, the models start to fail, on longer timescales $\gtrsim 1\, \text{Myr}$.
In order to obtain some clues as to what may be causing these differences, here we study their geometry, which reveals a preferential alignment that may point us towards the governing physical processes.

\subsection{The shape phase-space} \label{sec:shapes}

In Fig. \ref{fig:geometry_track}, we show the trajectories of the SNR \#22 for the different explosion models in the shape phase space, defined by the minor-to-major and semi-major-to-major axis ratio. 
By definition, at $t=0$ the SNR starts as a perfect sphere ($a/c = b/c=1$) and by deformation through various processes may evolve to become increasingly prolate ($b/c < 2/3$) or oblate ($a/c \rightarrow 0$ and $b/c > 2/3$).
In purple, we also show the trajectory of a
shearing sphere as described in the App. \ref{app:sphere_model}. The sphere of radius $r_{0} = 100\,\text{pc}$ is initially located at a galactocentric radius of $R_{0} = 8 \,\text{kpc}$ and is rotating at a constant rotation velocity, matching that measured in the simulation.

In the models N1 and N10, the trajectory in the shape phase-space is smooth, with almost constant minor-to-semi-major axis-ratio $a/b > 2/3$ and ever decreasing $a/c$, i.e. the SNRs are becoming increasingly prolate.
In model N1, $a/b$ is slightly larger than in N10, i.e. the SNR is slightly more prolate.
The simulated SNRs are significantly more deformed than the shearing sphere, which after 10 Myr is still quite spherical ($a/c \sim 0.75$, $b/c \sim 0.85$).

The trajectory in the model N1x10 has a kink, corresponding to the onset of further explosions, which deform the SNR in chaotic ways, ultimately leading to a more spherical shape. 
The final shape is similar to that of the shearing sphere.

In Fig. \ref{fig:snr_shapes} we show the 
locations of our sample of SNRs in the shape phase-space at various characteristic points in time. Markers are colored by the ambient density.
In the panel corresponding to the last snapshot at 10 Myr, we compare the shape of the LB derived from the data products of \citet{2024A&A...685A..82E}, shown as an orange and blue square, to our simulated sample.

The three panels corresponding to shell-formation, the end of the PDS phase and the onset of the implosion reveal that most SNRs remain close to spherical throughout the main stages of SNR evolution, with $a/b \gtrsim 2/3$ and $b/c \gtrsim 2/3$.
In contrast, SNRs in very low density ambient media $n_{\text{H}} \lesssim 10^{-2} \, \text{cm}^{-3}$ already begin to deviate from spherical symmetry before shell formation, likely due to their older age and larger size at the the same evolutionary stage, indicating that they are likely tracing a more anisotropic environment than their high-density counterparts.

After 10 Myr the trend is reversed. The SNRs with the highest ambient densities are deformed the most, exhibiting highly anisotropic shapes $a/c \sim 0.2$ with a wide range of geometries $1/3 \lesssim b/c \lesssim 1$.
Only 4 SNRs remain spherical, with most SNRs being slightly prolate and some oblates. 
The LB has a usual shape for an SNR with its ambient density, being slightly prolate with $a/b \sim 0.8$ and $a/c \sim 0.5$.

\subsection{Alignment of SNRs within the galaxy} \label{sec:directions}

\begin{figure}
 \includegraphics[width=\linewidth, clip=true]{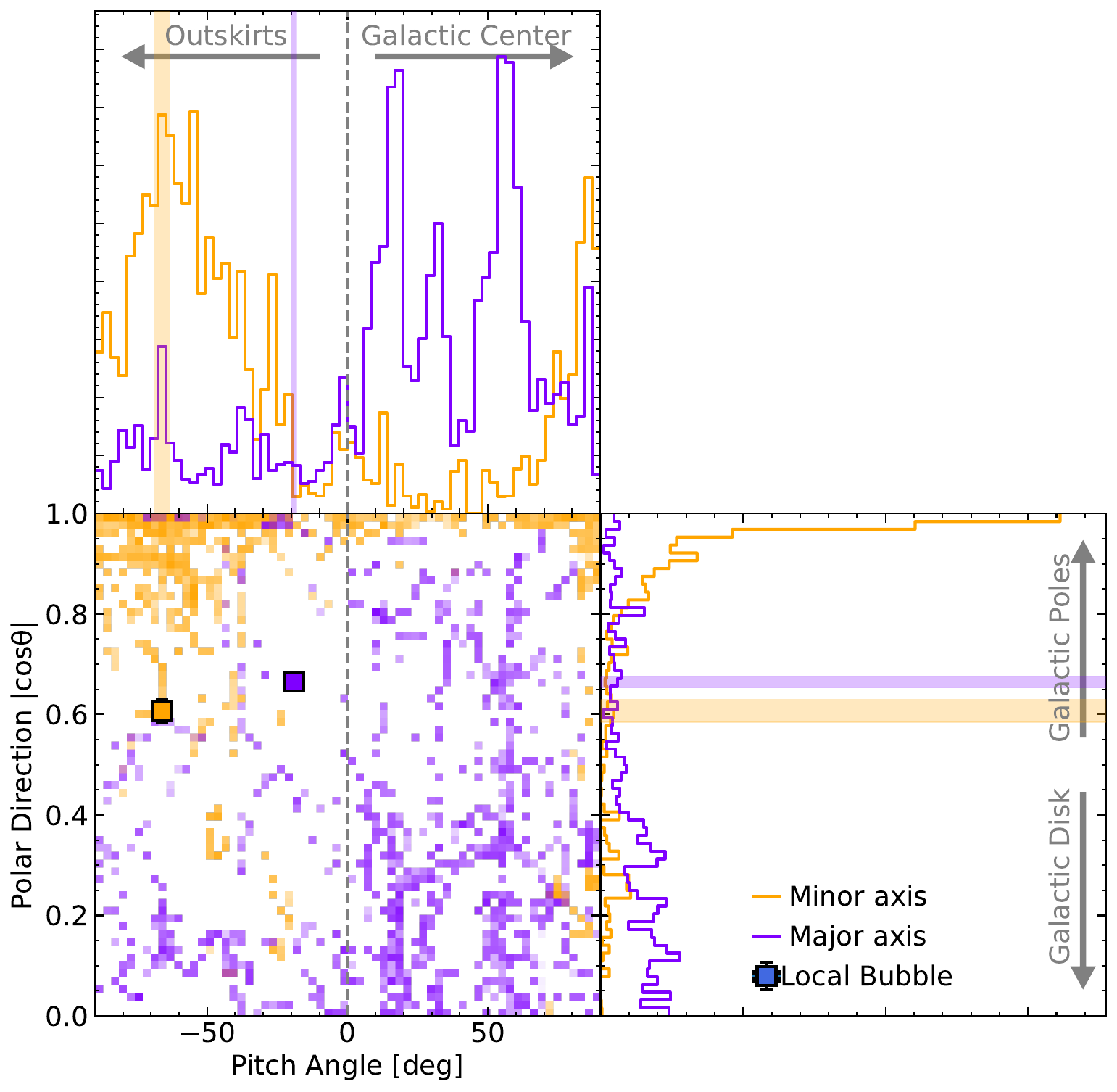}
 \caption{Time-span-weighted-histograms showing the distribution of SNR directions for asymmetric SNRs. Distributions of oblate (prolate) SNRs are colored orange (violet). For oblate (prolate) SNRs we show the direction of the minor (major) axis. 
 We show the polar direction normal to the disk plane and the pitch angle relative to the direction of galactic rotation. 
 We do not differentiate between directions above or below the disk.
 Positive pitch angles point between the galactic center and negative angles point towards the galactic outskirts.
Orange and purple squares depict the directions of the minor and major axes of the LB, respectively, derived from the 3D dust maps of \citet{2024A&A...685A..82E}. The $1\, \sigma$-ranges for the LB are also indicated as shaded areas in the one-dimensional histograms.
 Oblate SNRs tend to point vertically out of the disk and towards the galactic outskirts with a typical pitch angle of $\alpha_{\text{minor}} \lesssim -50 \, \text{deg}$. 
 On the other hand, prolate SNRs tend to lie within the disk plane 
 $\left|\text{cos}\left(\theta\right)\right| \lesssim 0.5$
 pointing slightly towards the galactic center $\alpha_{\text{major}} \sim 10 - 60 \, \text{deg}$.}  \label{fig:snr_directions}
\end{figure}

In the previous subsection we have shown that the simulated SNRs evolve towards increasingly anisotropic geometries, suggesting that they may expand more in certain directions than others. In order to check, whether there are any preferential directions, in Fig. \ref{fig:snr_directions} we show the time-span weighted distribution of the pitch angles and the magnitude of the polar directions as defined in Sect. \ref{sec:ellipsoid} of the minor axis (oblates, orange) and major axis (prolates, purple).
We also show the alignment of the minor- and major-axes of the LB derived from the data products of \citet{2024A&A...685A..82E}, shown as an orange and a purple square, respectively. The $1\,\sigma$-confidence intervals are shown as shaded regions in the one-dimensional histograms.

We find that for most of the time, the minor axis of the oblate SNRs is pointing perpendicular to the disk plane, with a broad distribution of negative pitch angles, centered around $\alpha_{\text{oblate}} \sim -60^{\circ}$.
In contrast, in the case of prolate SNRs, the polar direction of the major axis is broadly distributed, with most of the weight lying below $\left|\text{cos}\left(\theta\right)\right| \lesssim 0.5$, corresponding to the directions within the galactic plane. 
The distribution of pitch angles has three peaks around $\alpha_{\text{major}} \sim 15^\circ, \, 25^\circ \, \text{and} \, 50^\circ$, in line with the expectations for structures deformed by shear (App. \ref{app:sphere_model}, see also the alignment of underdense substructure in Fig. \ref{fig:disk_projection}).

While the LB is slightly prolate, its minor axis points in a direction in agreement with that of oblate SNRs.
On the other hand, its major axis is pointing slightly towards the galactic outskirts and is slightly more perpendicular than the bulk of our sample of SNRs.

\subsection{Deformation timescale}\label{sec:timescale}

\begin{figure}
 \includegraphics[width=\linewidth, clip=true]{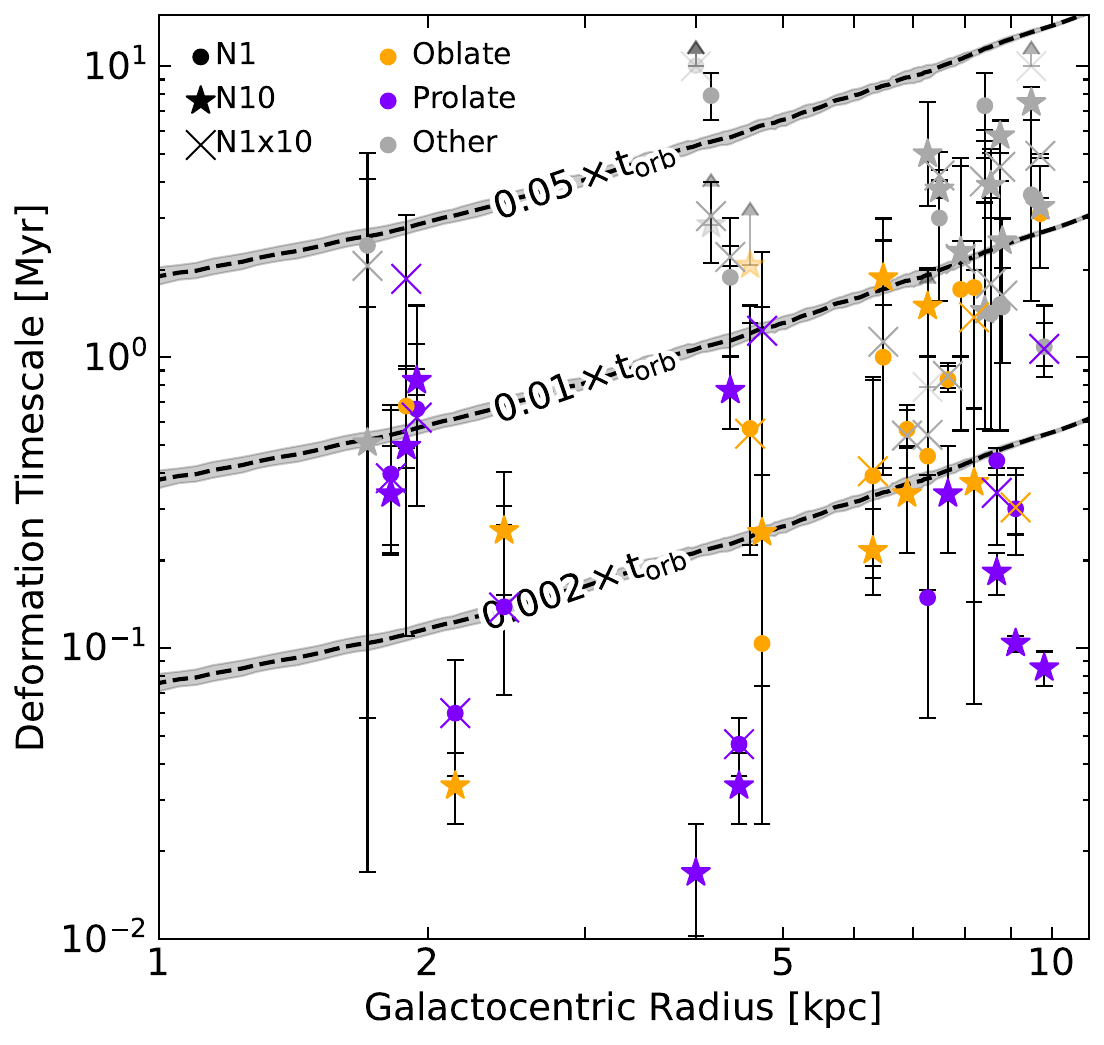}
 \caption{Deformation timescale as a function of galactocentric radius. 
 Uncertainties arise due to the finite spacing of the snapshots and due to the choice of $Z_{\text{ej, thr}}$.
 For visibility, markers are slightly shifted around their respective radii $R_{\text{gal}} = $ 2, 4.5 and 8 kpc, with the same shift used for the same SNR, but different explosion model. 
 Typical deformation timescales are on the order of a few percent of the orbital timescale at each radius, slightly shorter than what is expected from deformation by galactic shear alone.}  \label{fig:deformation_radius}
\end{figure}

\begin{figure}
 \includegraphics[width=\linewidth, clip=true]{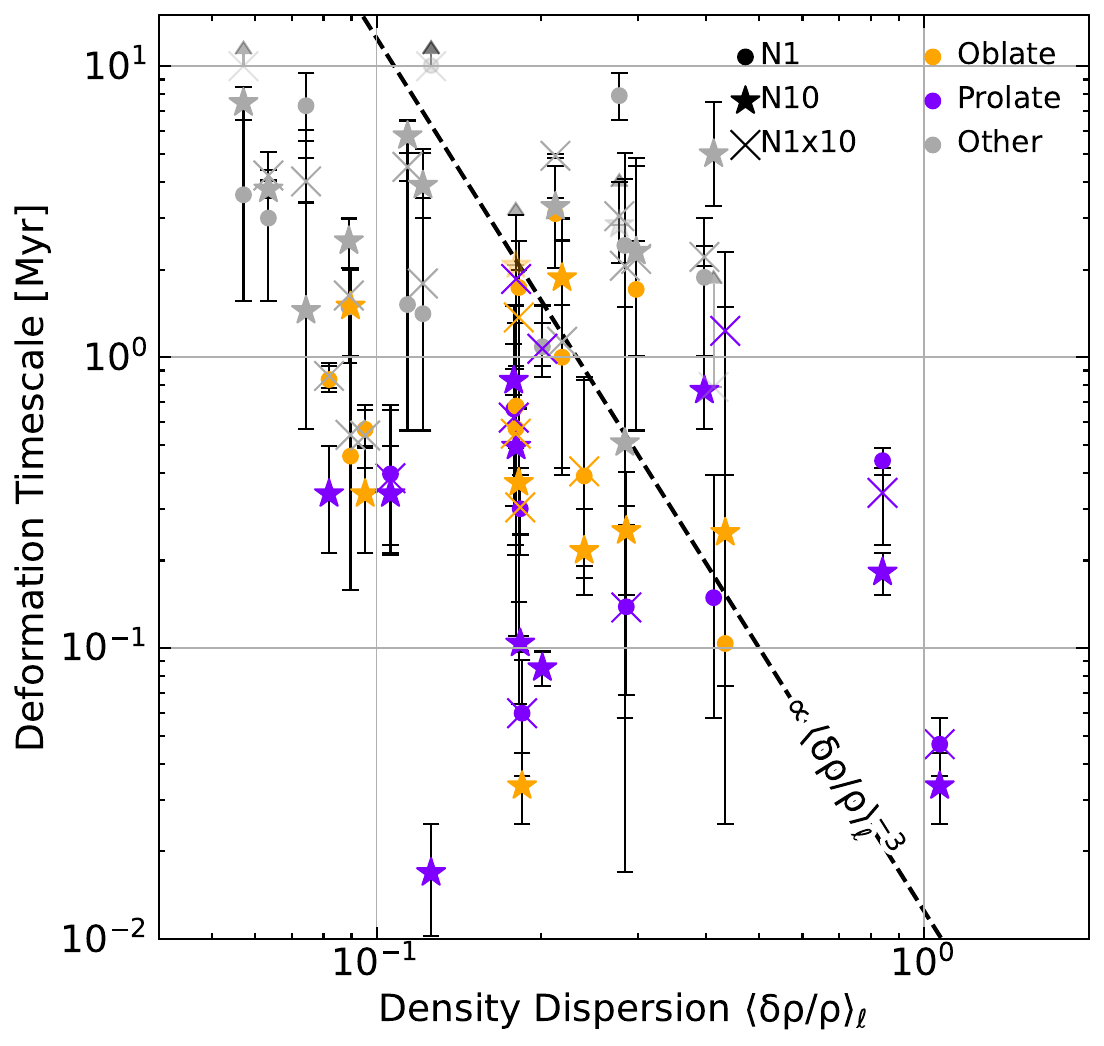}
 \caption{Deformation timescale as a function of density dispersion. 
 Uncertainties arise due to the finite spacing of the snapshots and due to the choice of $Z_{\text{ej, thr}}$.
 The deformation timescale is roughly $\propto (\delta\rho/\rho)^{-3}$ with significant scatter, qualitatively in line with the expectation that SNRs are deformed earlier in more anisotropic media.}  \label{fig:deformation_dispersion}
\end{figure}

In the previous subsections we have found, that our sample of simulated SNRs evolving into the shearing, stratified ISM of the SISSI galaxy grow increasingly anisotropic, assuming a geometry that aligns with the sheared structure of the galaxy.
The time it takes for an initially spherically symmetric structure such as SNR to become deformed hints at the processes governing the deformation.
To this end, we define the deformation timescale as the time at which the minor-to-major ratio drops below $a/c = 2/3$.

The shearing-sphere model (App. \ref{app:sphere_model}) indicates that shear can deform a spherical structure within a few percent of an orbital timescale.
To test, whether shear alone is enough to explain the deformation of the SNRs we show the deformation timescale as a function of galactocentric radius in Fig. \ref{fig:deformation_radius}. 
Markers are colored based on the shape classification of the SNRs.

The majority of the SNRs is deformed within $\lesssim 1\,\%$ of the orbital timescale, with several SNRs being deformed much before even a thousandth of an orbit.
More spherical SNRs, i.e. SNRs that are classified neither as prolate or oblate, tend to have longer deformation timescales, more plausibly explicable by shear alone.
There are relatively more oblate SNRs at larger galactocentric radius.
Overall, the deformation of the SNRs is too rapid to be explained by shear alone.

Another likely relevant source of deformation are preexisting density anisotropies in the ambient ISM, which imprint onto the geometry of the SNRs as they expand into them \citep{2023MNRAS.523.1421M}.
We quantify the degree of spatial variation in the density field, by measuring its relative variation at $t=0$ across spatial scales, by averaging over nested ISM patches of side-lengths $\ell = 0.2$, 0.5, 1, 1.5 and 2 kpc, corresponding to the scatter in Fig. \ref{fig:cornerplot_ICs}.

In Fig. \ref{fig:deformation_dispersion} we show the deformation timescale as a function of the thus defined density dispersion.
We find a steep decline in the deformation timescale with increasing density dispersion $\propto (\delta\rho/\rho)^{-3}$, in qualitative agreement with the expectation that indeed anisotropies in the density distribution might be dictating the geometry of SNRs in a turbulent ISM.
Since the dense structures in the ISM themselves are subject to differential rotation, they can be stretched out considerably by galactic shear over timescales that are much longer than the age of the SNR and thus imprint a relatively larger degree of anisotropy than the expansion of an SNR subject to shear alone.

\section{Discussion} \label{sec:discussion}

In the previous sections we have described the evolution of the geometry of SNRs expanding into an ISM structured by the complex interplay of gravity, galactic rotation and turbulence.
In the following, we will discuss some of the limitations of our simulations, how our results compare to observations of SNRs as well as
implications of our findings to the study of galaxy evolution and the structure of the ISM.

\subsection{Limitations}\label{sec:limitations}

The SISSI simulation suite aims to simulate the evolution of SNRs in a realistic galactic environment.
Of course, simulating a realistic galactic ISM is challenging and the numerical prescriptions and sub-grid physics involved can greatly influence the phase structure and the morphology of the galaxy as a whole.
A discussion of the quality of the simulated ISM of the SISSI galaxy, which is part of the AVALON simulation suite, focusing on the modeling of various aspects of galaxy evolution related to the structure of the ISM is out of the scope of this work, and will be presented elsewhere (M. Behrendt et al. 2025, in prep.).

Highly resolved simulations of SNRs, alongside an entire galaxy are computationally challenging and the computational resources to resolve large patches of the ISM with our maximum zoom-in resolution, are out of reach for currently available computing hardware.
Therefore, we had to resort to the refinement strategy outlined in Sect. \ref{sec:zoom-in}, which may itself introduce numerical artifacts. 
In the case of a uniform medium and without gravity, \citet{2024ApJ...965..168R} have shown that our method of a co-evolving refinement region does not greatly affect the evolution of SNRs.
However, we do find some differences in the star-formation activity and the partition of energy between the models N0 and N0\_zoom, suggesting that the properties of the background ISM might differ substantially, based on the resolution.
We account for this fact, by relaxing the initial conditions for 50 kyr, however it remains unclear, whether there is an optimal relaxation duration, given that the differences between the high- and low-resolution ISM do not seem to reach an asymptotic state and instead may be attributed to chaos, due to unresolved gravitational collapse coupled to stochastic star-formation.

More detailed modelling of physical processes, such as cosmic rays, magnetic fields, thermal conductivity, non-equilibrium radiation chemistry as well as more detailed stellar models, that include sources of early stellar feedback can influence the dynamics as well as the geometry of SNRs \citep[e.g.][]{2019MNRAS.483.3647G, 2023MNRAS.523.1421M, 2025ApJ...980..167D, 2024arXiv241112809G}.
In the present work we opted for a lightweight physics model, to lower the computational cost, allowing for higher resolution.
However, future efforts involving more detailed physics models may certainly be worthwhile.

\subsection{Observations of SNR geometry}\label{sec:observations}

As SNRs evolve, so do the wavelengths of light in which they can be observed.
Young SNRs are usually observed in the optical, infrared and X-ray \citep[e.g.][]{2023ApJ...945L...4F, 2024ApJ...961...32K, 2024ApJ...976L...4D} and their geometry is dictated by the explosion mechanism as well as their immediate surroundings.
These SNRs are usually fairly close to spherical symmetry, justifying our spherically symmetric injection of energy and ejecta mass.

Once SNRs enter the ST phase, they are extremely hot and bright in X-rays with diffuse X-ray emission coming from their center \citep{2023MNRAS.521.5536K, 2024ApJ...962..179R}. 
In this evolutionary stage, most SNRs are very close to spherically symmetric, though interactions with nearby clouds can lead to asymmetric features \citep{2024ApJ...975L..28C} in agreement with our sample of simulated SNRs, where the majority of SNRs remain close to spherically symmetric, with the exception of those exploding in low-density conditions, which are likely affected by interactions with clouds and low-density channels.

Galactic SNRs are usually only observed until shortly after they enter the radiative stage, as they quickly become too faint to be observed.
Observed radiative SNRs, tend to be quite spherically symmetric \citep{2024MNRAS.52711685P}, with few exceptions due to interactions with nearby density structures \citep{2024A&A...684A.178A}, in agreement with our simulations, which indicate that most SNRs remain spherically symmetric between shell-formation and the end of the PDS phase.

Older SNRs are usually too faint to be observed directly. 
However, once they become large enough, they may be indirectly observed by looking for large cavities in the dust distribution.
Such large cavities are routinely observed both in the Galaxy \citep{2020A&A...636A..17P, 2022Natur.601..334Z, 2022MNRAS.516L..35L, 2023ApJ...953..145V} as well as in nearby galaxies \citep{2023ApJ...944L..24W, 2023MNRAS.524.4907S, 2024A&A...690A.161L}. 
At first glance, observed SBs exhibit a wide variety of shapes and orientations, however due to the scarcity of detailed analyses of their shapes and orientation with respect to galactic structure, it is difficult to say to what degree, the observed sample agrees or disagrees with our simulated sample.

Fortunately, due to the recent 3D dust map made available by \citet{2024A&A...685A..82E}, we are in a position to study the geometry of the LB, a SB believed to be excavated by the SN explosions of $\sim \mathcal{O}\left(10\right)$ massive stars within the last $\sim \mathcal{O}\left(10\right)\,\text{Myr}$ \citep{2006A&A...452L...1B, 2021Sci...372..742W, 2022Natur.601..334Z}.
We find that, while the orientation of the LB is slightly unusual for an SB its age, its shape fits right into the range of shapes that we report for our sample of simulated SNRs after 10 Myr.
Moreover, we find that its effective size of $R_{\text{eff}} \sim 212.3 \pm 1.0 \text{pc}$ is on the lower end of sizes obtained after 10 Myr.
In a companion paper focusing on this aspect (L. Romano et al. 2025) we further discuss the implications, in particularly regarding current age estimates of the LB.

\subsection{Implications and future directions}\label{sec:implications}

Our numerical simulations show, that the geometry of evolved SNRs is changed due to the complex interplay of the expanding shell and a variety of environmental factors.
Since different processes affect the geometry of the SNR on different timescales, it might be possible to disentangle their contributions and deepen our understanding of the underlying physical processes shaping the internal structure of galaxies.

By leveraging novel analysis techniques \citep{2024A&A...685A..82E}, and increasingly detailed observations \citep{2023A&A...674A...1G} it will soon be possible to study the geometry of an ever growing sample of galactic SNRs.
Already, the data products of \citet{2024A&A...685A..82E} can be used to study the neighboring known SBs, such as the Per-Tau SB \citep{2021ApJ...919L...5B} and GSH 238+00+09 \citep{1998ApJ...498..689H}, which could provide additional hints to the assembly of structures in the solar neighborhood.

While we have focused on SNR geometry in this work, there are many more aspects of SNR phenomenology that can be addressed by the SISSI simulations. 
In future studies we aim to investigate the role of SNRs in driving interstellar turbulence, their coupling to and potential driving of galactic outflows as well as triggered star-formation.

\section{Concluding Remarks}\label{sec:conclusion}

We have introduced the SISSI simulation suite, featuring 3D hydrodynamic zoom-in simulations of SNRs embedded in the realistic, self-consistently generated ISM of an isolated, Milky-Way-like galaxy, in order to deepen our understanding of various aspects of SNR physics.
In this work, we focus on the geometry of the SNRs and show how it can be used as a useful observational diagnostic for understanding the various environmental effects, affecting the evolution of the system.
Here we summarize our most important findings:
\begin{enumerate}
    \item The dynamics of young SNRs ($\lesssim 1\,\text{Myr}$) are well described by standard analytical models. However, these models become less accurate for SNRs exploding in low-density ($\lesssim0.1\,\text{cm}^{-3}$) environments, likely due to the large size of the SNRs, which increases the likelihood of interactions with both high- and low-density structures, such as clouds and channels.
    At later times, environmental effects start to dominate the dynamics of all simulated SNRs, leading to $\gtrsim 2$ times larger SNRs than predicted by models of SNRs in uniform, stationary media (Fig. \ref{fig:sizes}).
    \item SNRs tend to be deformed greatly on a timescale shorter than a few per cent of the orbital timescale, with SNRs in environments with larger density fluctuations being deformed earlier,
    as they tend to follow the geometry of dense structures, which are deformed by shear and aligned with the galactic rotation.
    \item The deformation of SNRs has preferred directions. The minor axis of oblate SNRs tends to be aligned with the galactic poles, with a slight tilt towards the galactic outskirts,
    suggesting that the vertical expansion of these SNRs is stalled by the graviational pull of the galactic disk.
    The polar angle of the major directions is broadly distributed, slightly favoring directions in the galactic plane, with pitch angles peaked between $\sim 15^\circ$ and $\sim 50^\circ$ slightly pointing towards the galactic center,
    in agreement with the expectation from alignment due to galactic shear.
    \item The LB has a typical geometry for a SB of its age and size, however it appears slightly small compared to the size of the SNRs in the SISSI sample at 10 Myr.
    This suggests, that previous estimates of the age, based on idealized, one-dimensional models of SB expansion might need to be revised. The LB might be just old -- and large --  enough to be affected by its galactic environment, which makes it a unique laboratory to study the expansion of SBs at the interface between local ISM physics and galactic dynamics.
\end{enumerate}
We conclude that SNR geometry offers a novel observational tool for understanding the complex physics of galaxies and their impact on galactic substructure, leveraging the full potential of recent high quality observations.

\begin{acknowledgements}
We thank the anonymous referee for their insightful comments and suggestions that helped to improve the quality of this work.
Computations were performed on the HPC systems Cobra and Viper at the Max Planck Computing and Data Facility.
This research was funded by the Deutsche Forschungsgemeinschaft (DFG, German Research Foundation) under Germany's Excellence Strategy – EXC 2094 – 390783311.
LR thanks the developers of the following software and
packages that were used in this work: \textsc{Julia} v1.10.0 \citep{Julia-2017},
\textsc{Matplotlib} v3.5.1 \citep{Hunter:2007},
\textsc{Mera} v1.4.4 \citep{behrendt_2023_7934894},
\textsc{Ramses} v19.10 \citep{2002A&A...385..337T}, and
\textsc{Healpix} v2.3.0 \citep{2021ascl.soft09028T}
\end{acknowledgements}

\bibliographystyle{aa} 
\bibliography{bibliography}

\appendix

\section{The ISM of the SISSI galaxy}\label{app:ISM}
\begin{figure}
 \includegraphics[width=\linewidth, clip=true]{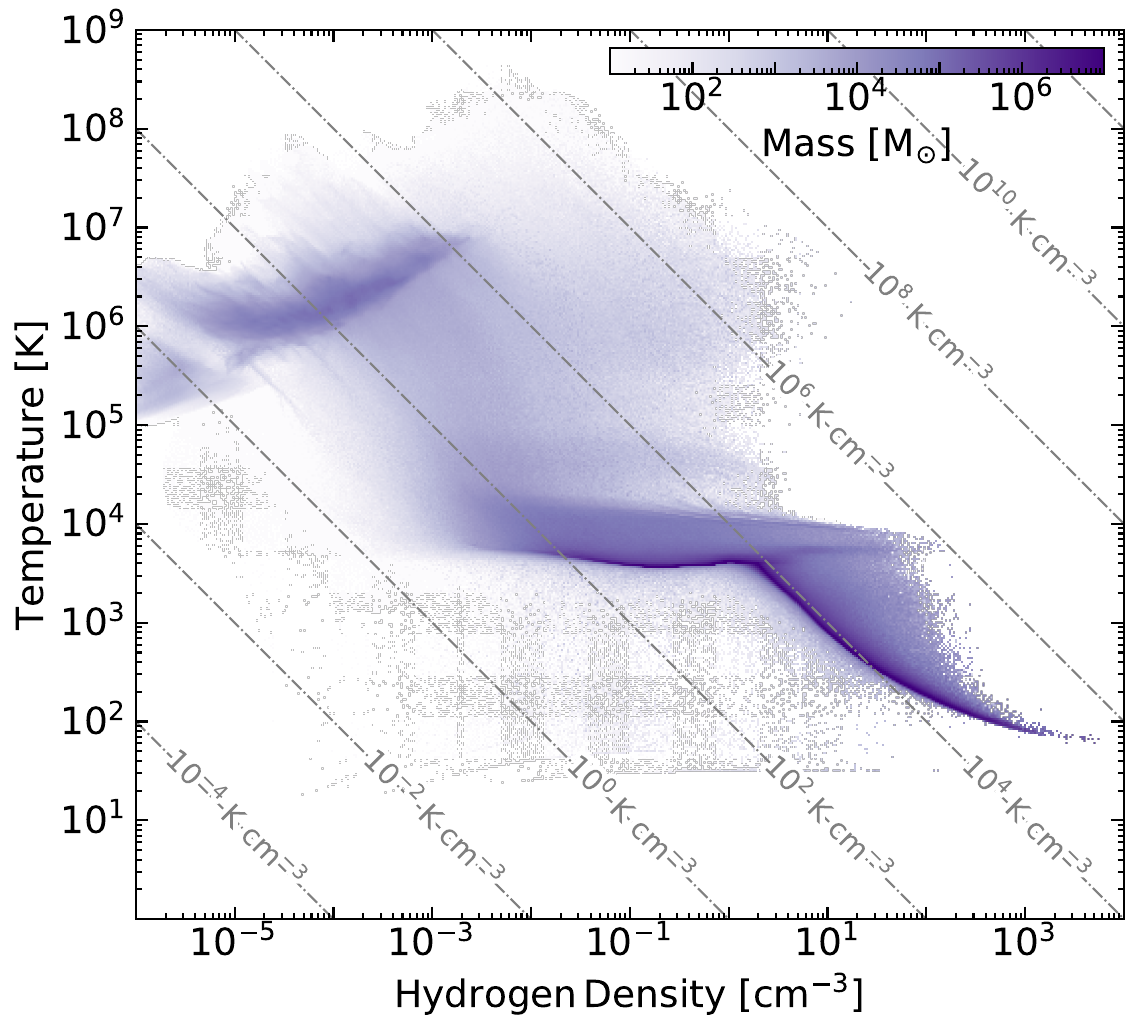}
 \caption{Temperature-density phase-diagram of the global ISM. The galactic ISM features a stable phase at $P/k_{\text{B}}\sim 10^4\,\text{K cm}^{-3}$ for $n_{\text{H}} \sim 1 - 100\,\text{cm}^{-3}$, and an unstable phase at $T \lesssim 10^4\, \text{K}$, which becomes stable below $n_{\text{H}} \sim 1\,\text{cm}^{-3}$.} 
 \label{fig:phase_diagram}
\end{figure}
\begin{figure*}
 \includegraphics[width=\linewidth, clip=true]{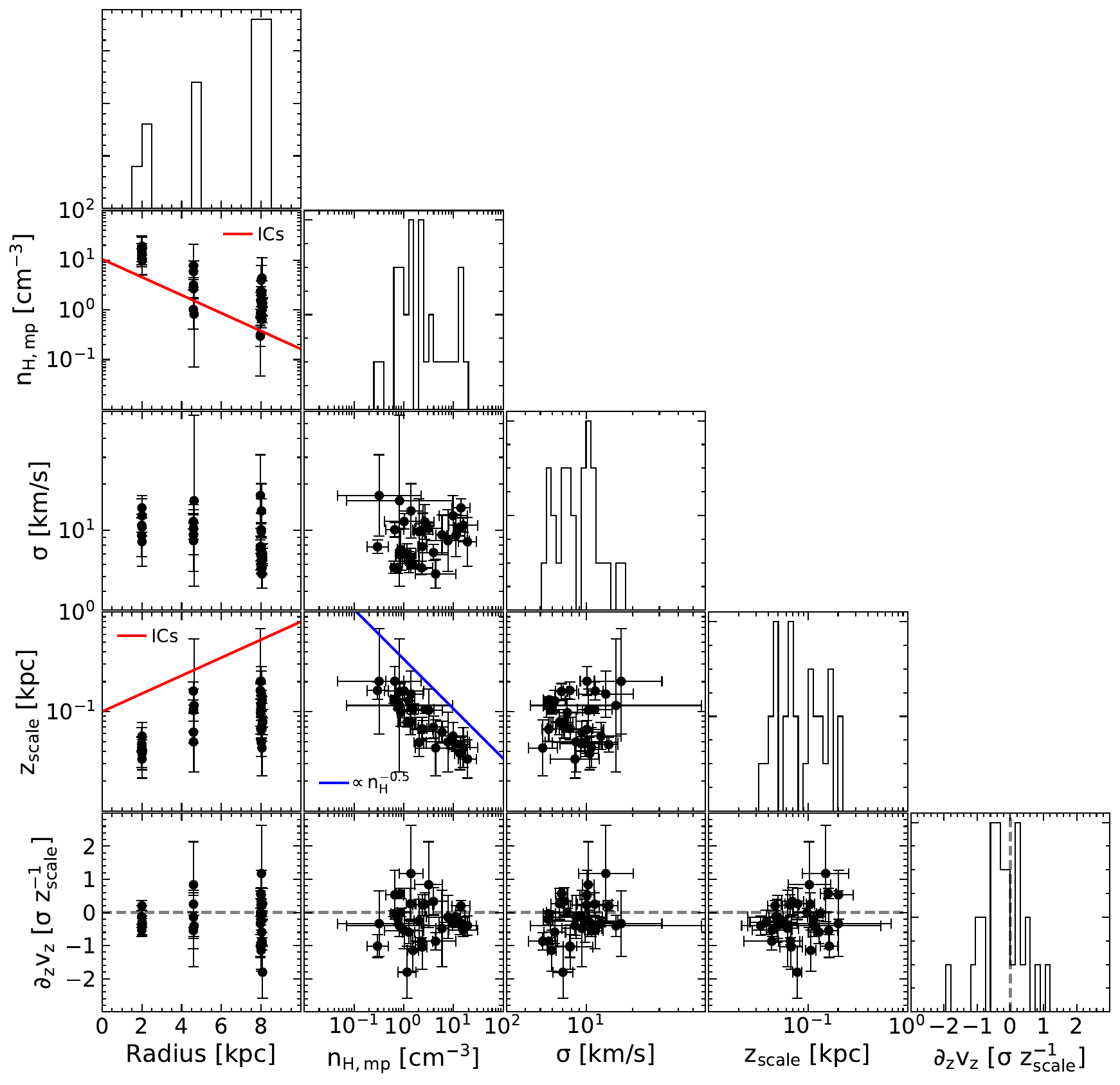}
 \caption{Corner plot showing the distribution of ISM properties (Galactocentric radius $R$, midplane density $n_{\text{H, mid}}$, velocity dispersion $\sigma$, vertical scale height $z_{\text{scale}}$ and vertical velocity gradient $\partial_{z} v_{z}$) at the SNR locations averaged over various length scales. 
 We average over quadratic apertures with side lengths $L = 0.2, \, 0.5, \, 1.0,\, 1.5 \, \text{and}\, 2.0\, \text{kpc}$. 
 As expected the mean of $n_{\text{H, mid}}$ decreases with $R$, with large scatter at $R > 2 \,\text{kpc}$. 
 Compared to the ICs, the mean density profile has steepened. 
 The velocity dispersion is roughly constant $\sigma \sim 10\, \text{km/s}$ throughout the disk with considerable scatter. 
 The disk scale height follows the scaling behavior predicted by vertical hydrostatic equilibrium of the gas $z_{\text{scale}} \propto \sigma / n_{\text{H}}^{0.5}$, but with a slightly lower normalization, likely due to the presence of stars.
 The vertical velocity gradient on average is zero with increasing scatter towards larger $R$. 
 The maximum values reveal a preferred scale of $\left|\partial_{z} v_{z}\right| \sim \sigma / z_{\text{scale}} \sim t_{\text{ff}}^{-1}$ indicating that these expansion and contraction motions correspond to gravitational breathing modes of the disk.
 The velocity gradients peak around $n_{\text{H, mid}} \sim 1 \, \text{cm}^{-3}$, i.e. the breathing of the disk coincides with the presence of multi-phase gas (see fig:\ref{fig:phase_diagram}).} 
 \label{fig:cornerplot_ICs}
\end{figure*}

The SNRs of the SISSI simulations expand into a complex environment which differs between different regions of the ISM.
In order to help interpreting the role of the ISM, here we describe the properties of this environment.

Figure \ref{fig:phase_diagram} shows the $T-n_{\text{H}}$ phase diagram of the gas in the SISSI galaxy at $t=0$.
In the ISM most of the gas is concentrated in two distinct gas phases: Warm neutral gas at $T \sim 7\times10^3 \, \text{K}$ for densities in the range $\sim 10^{-2} - 10^{2}\,\text{cm}^{-3}$, and colder gas at a constant pressure of $P \sim 10^{4} \, k_{\text{B}} \,\text{K}\,\text{cm}^{-3}$ for denser gas above $n_{\text{H}} \sim 1 \, \text{cm}^{-3}$.
For gas above $n_{\text{H}} \sim 1 \, \text{cm}^{-3}$ the warm phase is unstable and cools after $\sim 1\, \text{Myr}$.

Cold ($T \lesssim 10^2 \, \text{K}$) and dense ($n_{\text{H}} \gtrsim 10^2 \, \text{cm}^{-3}$) gas is star-forming and thus steadily being consumed.

The galactic ISM is surrounded by a hot, diffuse circumgalactic medium (CGM), corresponding to a roughly adiabatic phase with $T_{\text{CGM}} \sim 10^{7} \, \left(n_{\text{H}} / \left(10^{-3} \, \text{cm}^{-3}\right)\right)^{\gamma-1}$, with a maximum density of $\sim 10^{-3}\, \text{cm}^{-3}$ at a temperature of $\sim 10^{7}\, \text{K}$, in pressure balance with the cold ISM.

Of course, locally the ISM properties may not adhere to this simple picture. 
In Fig. \ref{fig:cornerplot_ICs} we show the initial properties of the local ISM at the various explosion sites.
All quantities are derived from density, and vertical velocity profiles averaged over ISM patches of side lengths $\ell = 0.2, 0.5, 1, 1.5 \, \text{and}\, 2 \, \text{kpc}$ parallel to the galactic plane centered around the explosion site.
The error bars indicate how much a quantity varies with scale. 

We define the vertical scale height as half the distance between the $\sim 12 \, \%$ and the $\sim 88\, \%$ mass percentiles of the density profile, roughly matching the definition of a $\text{sech}^{2}\left(z / z_{\text{scale}}\right)$-profile , as would be appropriate for a single-component, isothermal disk. While the simulated galaxy, is neither consisting of only a single component, nor is it isothermal, our definition still yields a reasonable definition for an effective gas scale-height.
Correspondingly, we define the midplane density as
\begin{equation}
    n_{\text{H, mp}} = \frac{1}{2 \,\text{tanh}\left(1\right)z_{\text{scale}}} \int_{z_{0}-z_{\text{scale}}}^{z_{0}+z_{\text{scale}}} n_{\text{H}}\left(z\right) \text{d}z  ~,
\end{equation}
where $z_0$ is the midplane, lying right in between the $\sim 12 \, \%$ and the $\sim 88\, \%$ mass percentiles of the density profile.

The velocity dispersion is defined as the average of the three components of the velocity dispersion vector, i.e. $\sigma^2 = \left(\sum_{i \in \left\{x,y,z\right\}} \sigma^2_{i}\right)/3$, within all vertical bins within the midplane, i.e. within $z_{0}\pm z_{\text{scale}}$.

We find a diverse range of midplane densities spanning over two orders of magnitude. The densities roughly follow the radial trend of the initial conditions, albeit with considerable scatter and a steepening towards $R_{\text{gal}} = 2\,\text{kpc}$. 
We thus expect SNRs at larger galactic radii to grow bigger, with more variation between regions. 

The velocity dispersion is roughly constant throughout the sample of regions with a typical value of $\sigma \sim 10\,\text{km s}^{-1}$, though with larger spatial variations in some regions.
SNRs should therefore merge with the ISM at around the same time in all regions.

The measured scale heights indicate that, while the overall trend follows the expected scaling from dynamical equilibrium considerations in a single-component disk, it is more compact due to the dominant gravitational potential of the stellar disk. 
SNRs will start to be affected by vertical stratification once their size grows similar to this scale height, indicating that these effects might become important earlier for SNRs in higher-density regions.

In some regions we find that the mean vertical velocity is increasing (decreasing) linearly as a function of height with a midplane vertical velocity gradient on the order of $\sigma / z_{\text{scale}}$. 
These motions appear to be strongest around $n_{\text{H, mp}} \sim 1 \text{cm}^{-3}$, indicating that gas at this density can be thermally unstable and is driven towards lower (higher) densities ($\partial_{z}v_{z} > 0 \, (<0)$).
We interpret these motions as disk \textit{breathing}-modes around the dynamical equilibrium, with an expected period of about a free-fall timescale, much longer than the dynamical timescale of an expanding SNR.
Thus we expect the velocity gradients to be frozen-in during the lifetime of an SNR.
SNRs expanding into a positive velocity gradient will grow faster as they sweep-up co-expanding material, while SNRs expanding into a collapsing region will be slowed down.

\section{Analytic theory of SNR evolution in a uniform medium}\label{app:theory}

In this section we briefly review the analytic theory for the dynamics of radiative SNRs and SBs \citep[see e.g.][]{2015ApJ...802...99K, 2022ApJS..262....9O, 2024ApJ...965..168R}.
We consider the case of spherical expansion driven by point-explosions with explosion energy $E_{\text{SN}} = 10^{51}\,E_{51} \,\text{erg}$ into a uniform medium with hydrogen number density $n_{\text{H}} = n_{0}\,\text{cm}^{-3}$, solar metallicity and pressure $P = \mu\,n_{\text{H}}\,\sigma^2$, where $\mu = 1.4$ is the mean atomic weight and $\sigma = 10\,\sigma_{1}\,\text{km s}^{-1}$ is the sound speed, which in a supersonically turbulent medium such as the ISM, may be replaced with the turbulent velocity dispersion.

Since here we are mostly interested in the dynamics of old radiative SNRs we skip the dynamics of the initial ejecta dominated expansion and start directly with that of adiabatic expansion; the so-called Sedov-Taylor (ST) phase \citep{1959sdmm.book.....S}.
The internal structure of the Sedov-Taylor blastwave is described by a similarity solution with similarity parameter $\xi = r / \left(E_{\text{SN}} t^3 / \rho\right)^{1/5}$, with $\xi_0 \approx 1.15167$ at the position of shock radius.
During the ST phase, the radially outward momentum increases as a function of time and is given by \citep{2015ApJ...802...99K}
\begin{eqnarray}
    p_{\text{ST}} = 2.21 \times 10^4 \, E_{51}^{4/5} \, n_{0}^{1/5} \, t_{3}^{3/5} \, \text{M}_{\odot} \, \text{km s}^{-1}~,
\end{eqnarray}
where $t = t_{3} \, \text{kyr} = t_{6}\, \text{Myr}$.

The ST phase ends, once radiative cooling becomes dominant and a thin shell forms right behind the shock front, after \citep{2015ApJ...802...99K}
\begin{equation}\label{eq:t_sf}
    t_{\text{sf}} \sim 44 \, E_{51}^{0.22}\,n_{0}^{-0.55}\,\text{kyr} ~,
\end{equation}
at which point the SNR has a size of
\begin{equation}\label{eq:r_sf}
    R_{\text{sf}} = 22.6 \, E_{51}^{0.29}\,n_{0}^{-0.42}\,\text{pc} ~,
\end{equation}
and a momentum of
\begin{equation}\label{eq:p_sf}
    p_{\text{sf}} = 2.17 \times 10^5 \, E_{51}^{0.93}\,n_{0}^{-0.13}\, \text{M}_{\odot} \, \text{km s}^{-1}~.
\end{equation}

Right after shell formation, the interior of the SNR is still hot
and at a higher pressure than the shell, which has a temperature of about $T_{\text{shell}} \sim 10^{4}\,\text{K}$ and is highly compressed relative to the ambient medium $\chi \sim 10$.
During this so-called pressure-driven snowplow (PDS) stage, the SNR expands $R \propto t^{2/7}$, leading to a slight enhancement of the radial momentum $\propto t^{1/7}$.
The PDS ends when the pressure in the bubble becomes comparable to the pressure in the shell \citep{2024ApJ...965..168R} after
\begin{equation}\label{eq:t_PDS}
    t_{\text{PDS}} \sim 0.15 \, E_{51}^{0.27} \, n_{0}^{-0.44}\,\text{Myr} ~,
\end{equation}
corresponding to a size of
\begin{equation}\label{eq:r_PDS}
    R_{\text{PDS}} \sim 32.1 \, E_{51}^{0.3} \, n_{0}^{-0.39}\, \text{pc} ~.
\end{equation}
Depending on the details of radiative cooling and incorporation of mass from the bubble into the shell, the radially outward momentum of the SNR is boosted by up to about $\sim 50 \, \%$ during the PDS.

Once the pressure in the interior of the bubble has dropped, the SNR expands solely due to its inertia $R \propto t^{1/4}$.
During this so-called momentum-conserving snowplow (MCS) phase, the pressure of the shell is proportional to the shock velocity $P_{\text{shell}} \propto t^{-3/2}$.
During this stage, the back of the shell is unstable and a reflected shockwave or implosion is driven into the interior of the SNR once the pressure of the shell becomes comparable to that of the ambient medium \citep{2024ApJ...965..168R} after
\begin{equation}\label{eq:t_launch}
    t_{\text{launch}} \sim 0.5 \, E_{51}^{0.27}\, n_{0}^{-0.44} \, \sigma_{1}^{-4/3}\, \text{Myr} ~,
\end{equation}
at which point the SNR has a size of 
\begin{equation}\label{eq:r_launch}
    R_{\text{launch}} \sim 43.6 \, E_{51}^{0.3}\,n_{0}^{-0.39} \, \sigma_{1}^{-1/3}\, \text{pc} ~.
\end{equation}
These expressions differ from those derived by \citet{2024ApJ...965..168R}, due to the differences in our model for the ambient pressure.

There are various different models for the case of an SB, driven by subsequent SN explosions \citep[e.g.][]{2019MNRAS.490.1961E, 2022ApJS..262....9O}. The basic assumption in these models is that, if the age of the SB is greater than the average time between SN explosions $t \gg \Delta t_{\text{SN}} = \Delta t_{6} \, \text{Myr}$, the expansion can be approximately described by that of a wind with a constant mechanical luminosity $L = E_{\text{SN}} / \Delta t_{\text{SN}} = 10^{45} \, L_{45} \, \text{erg yr}^{-1}$.
The dynamics of a radiative SB depend on the efficiency of energy dissipiation, e.g. due to radiative cooling, facilitated by thermal conduction and turbulent mixing of the hot interior and the cold shell.
When energy injection dominates over dissipation, the expansion can be described by that of an energy-driven wind \citep{1977ApJ...218..377W, 2019MNRAS.490.1961E}.
In contrast, if cooling losses dominate, the expansion is effectively momentum-driven \citep{2021ApJ...914...90L, 2022ApJS..262....9O, 2024ApJ...970...18L}.

In the model N1x10, $\Delta t_{\text{SN}} = 1 \, \text{Myr} \gg t_{\text{cool}}$, corresponding to the momentum-driven regime.
The size of a momentum-driven SB is given by Eq. 26 of \citet{2022ApJS..262....9O}
\begin{equation}\label{eq:momentum-driven-wind}
    R = 40 \, t_{6}^{1/2} \, L_{45}^{0.23}\,n_{0}^{-0.28}\,\text{pc} ~.
\end{equation}

\section{Shearing-sphere model}\label{app:sphere_model}
\begin{figure}
 \includegraphics[width=\linewidth, clip=true]{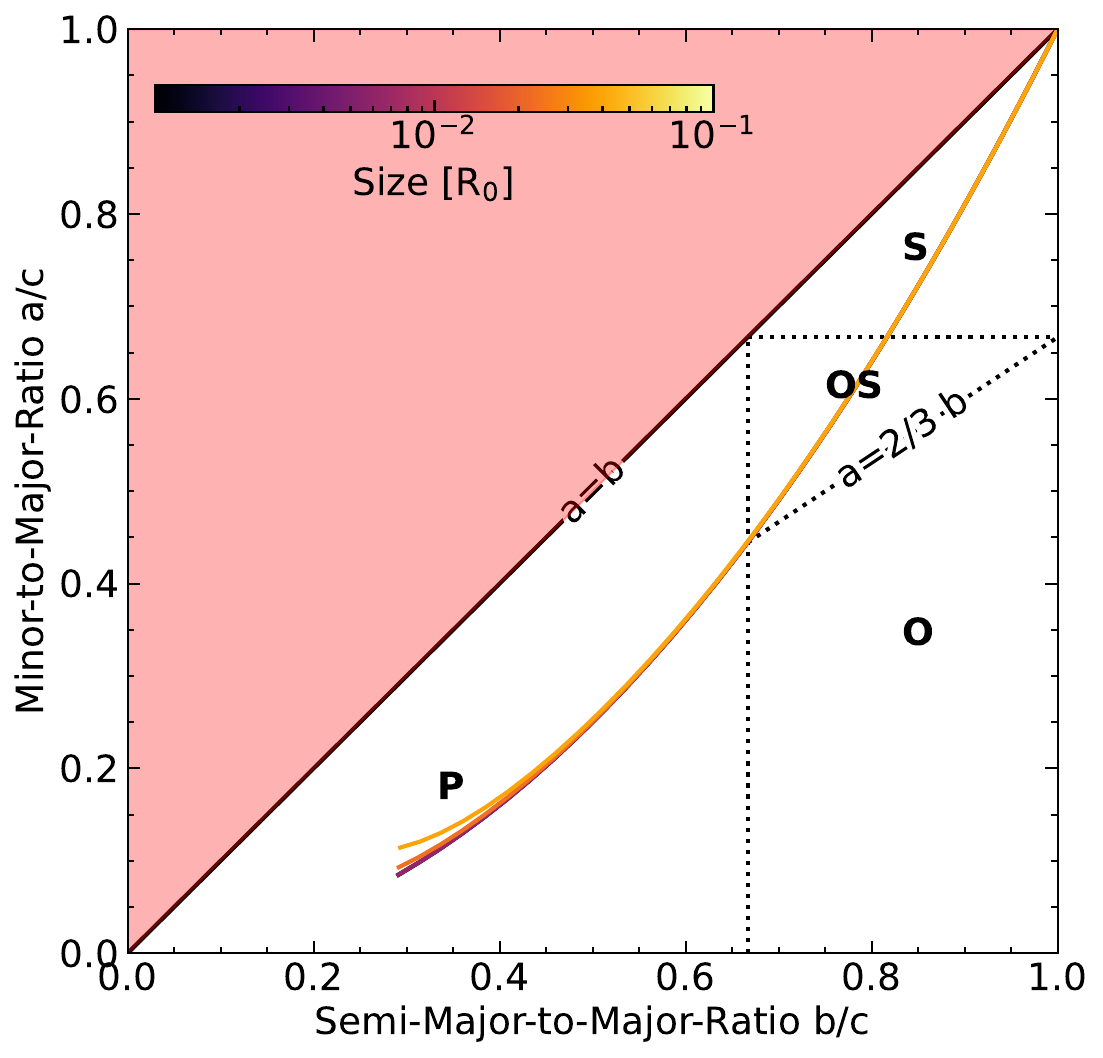}
 \caption{Same as Fig. \ref{fig:geometry_track} for the shearing sphere model for spheres with different sizes, evolved for $0.5 \, t_{\text{orb}}\left(R_{0} - r_{9}\right)$.
 The phase-space trajectories of the spheres with different sizes are almost identical.} 
 \label{fig:geometry_track_model}
\end{figure}
\begin{figure}
 \includegraphics[width=\linewidth, clip=true]{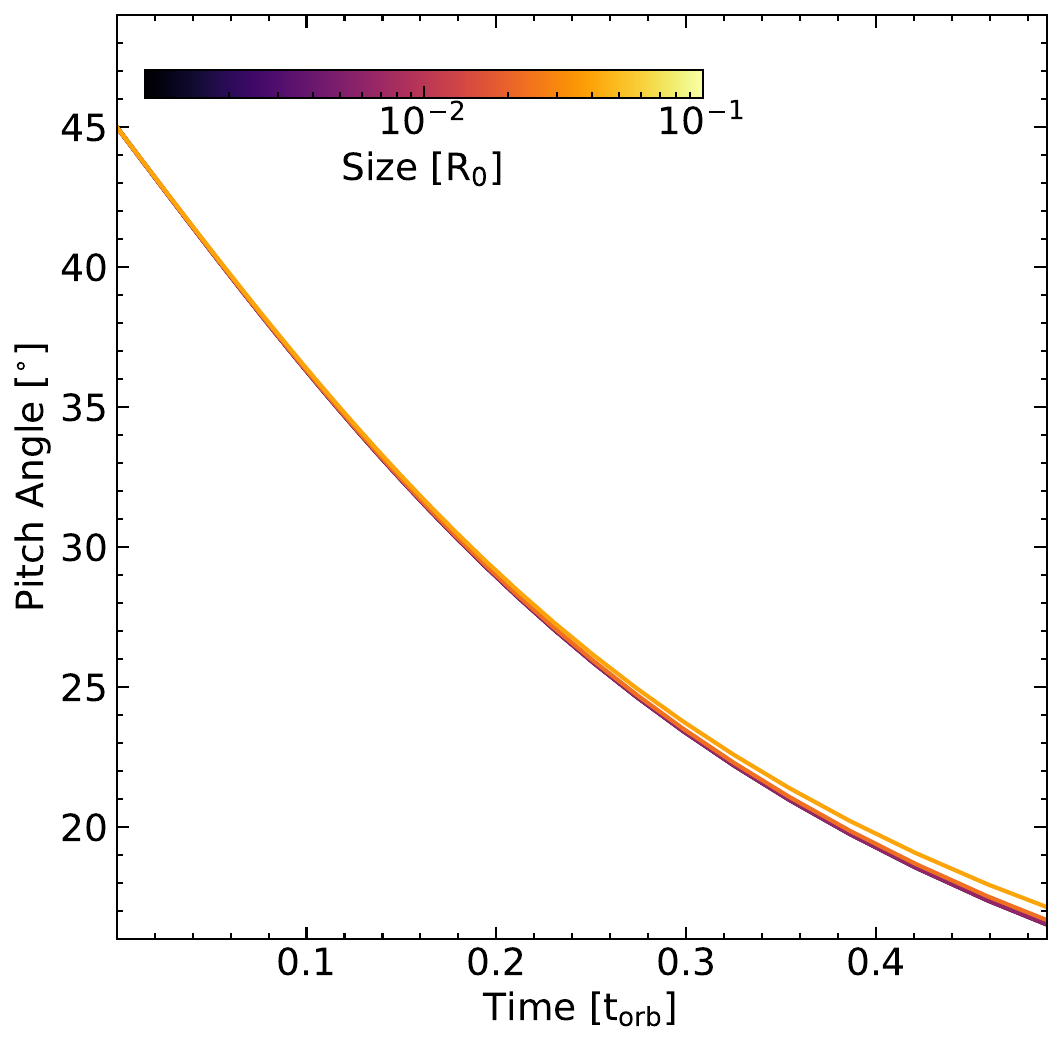}
 \caption{Time evolution of the pitch angle of the major axis of the shearing sphere for spheres with different sizes.
 The pitch angle starts off near $45\,^{\circ}$ and decays over time. Larger spheres tend to have slightly larger pitch angles.} 
 \label{fig:pitch_angles_model}
\end{figure}
\begin{figure}
 \includegraphics[width=\linewidth, clip=true]{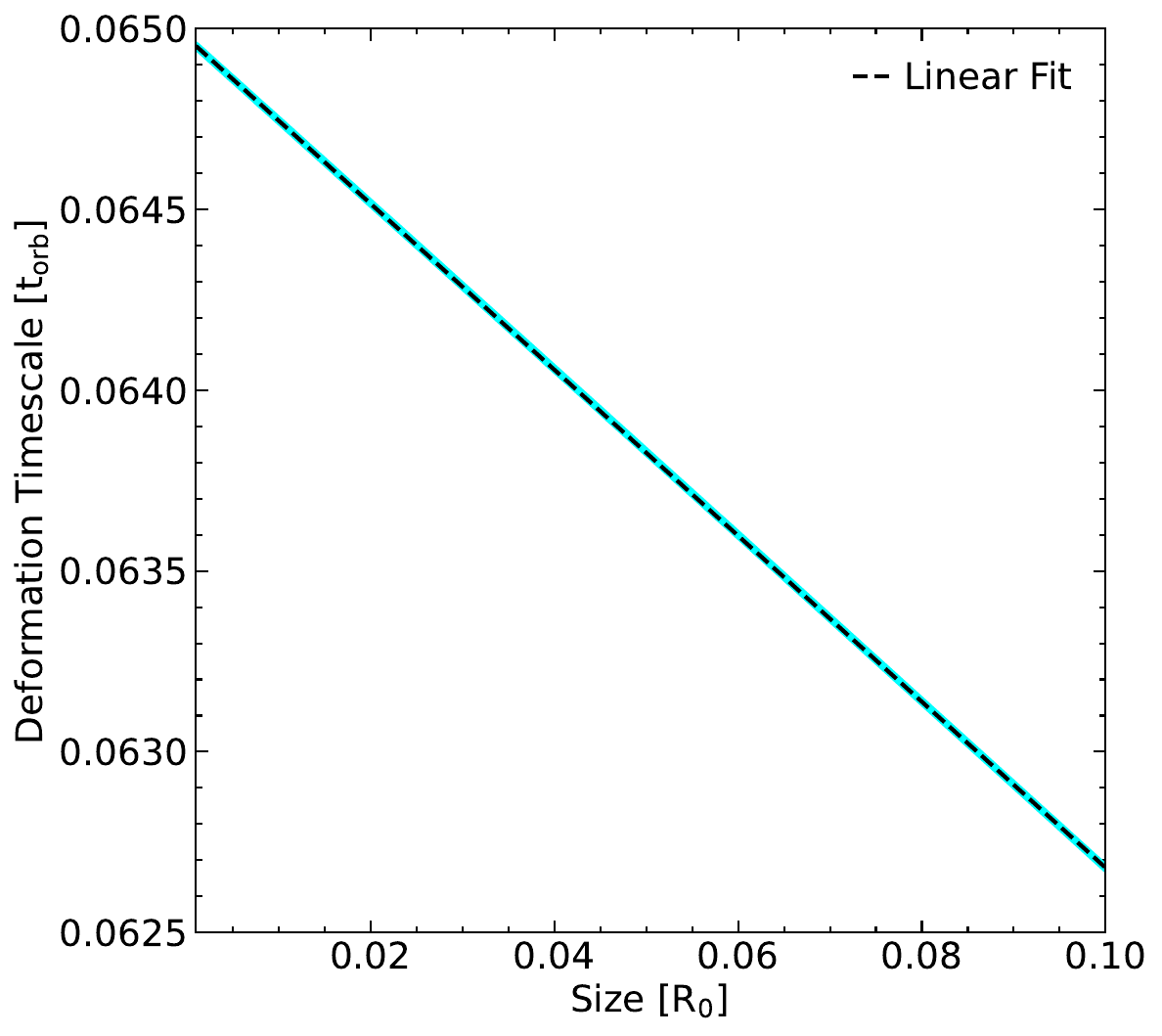}
 \caption{Deformation timescale as a function of size. 
 Spheres are deformed greatly after $\sim 6.5$ per cent of $ t_{\text{orb}}\left(\bar{R}\right)$, where $\bar{R} = pR_{0} + (1-p)\left(R_{0} - r_{0}\right)$ is a characteristic radius lying between $R_{0} - r_{0}$ and $R_{0}$. Our linear fit yields $p\sim0.65$.} 
 \label{fig:deformation_timescale_model}
\end{figure}

In order to model the deformation by shear, here we derive the simplest possible model: starting from $t=0$, a sphere of radius $r_0$, centered in the galactic midplane at a galactocentric radius $R_{0}$, is subjected to differential rotation with a constant rotation speed $V_{\text{rot}}$, corresponding to an angular frequency of $\Omega\left(R\right) = V_{\text{rot}} / R$. Correspondingly, the orbital timescale is $t_{\text{orb}}\left(R\right) = 2\pi \,\Omega^{-1}\left(R\right)$.

We parameterize the surface of the sphere using the polar angles $\theta$ and $\varphi$ at $t=0$
\begin{equation}
    \mathbf{\Phi}_{t}\left(\theta, \phi; r_{0}\right) = \begin{pmatrix} R \,\text{cos}\left(\phi_{0} + \Omega t\right) \\ R\,\text{sin}\left(\phi_{0} + \Omega t\right) \\ r_{0}\,\text{cos}\left(\theta\right)\end{pmatrix} ~,
\end{equation}
where
\begin{equation}
    R^2 = R_{0}^2 + 2R_{0} r_{0}\, \text{cos}\left(\varphi\right)\,\text{sin}\left(\theta\right) + r_{0}^{2}\,\text{sin}^{2}\left(\theta\right)
\end{equation}
is the galactocentric radius of the point on the surface of the sphere
and the initial azimuthal angle $\phi_0$ is defined by
\begin{eqnarray}
    R \, \text{cos}\left(\phi_{0}\right) &=& R_0 + r_0\,\text{sin}\left(\theta\right)\,\text{cos}\left(\varphi\right) \\
    R \, \text{sin}\left(\phi_{0}\right) &=& r_0\,\text{sin}\left(\theta\right)\,\text{sin}\left(\varphi\right) ~.
\end{eqnarray}
Due to the differential rotation, parts of the sphere that are at a larger galactocentric radius lag behind and the parts that are at a smaller $R$ advance ahead, leading to deformation.
It can be shown that in spite of the deformation, the volume remains constant.

We measure the geometry as defined, in Sect. \ref{sec:ellipsoid}, i.e. by computing the shape tensor
\begin{equation}
    S_{ij} = V^{-1}\int_{V} \left(\left\lVert \Delta\mathbf{\Phi}_{t}\right\rVert^2\delta_{ij} - \Delta\Phi_{t, i}\Delta\Phi_{t, j}\right)\text{d}V ~,
\end{equation}
where $\Delta\mathbf{\Phi}_{t} = \mathbf{\Phi}_{t}\left(\theta, \phi; r\right) - \mathbf{\Phi}_{c}\left(t\right)$ is the coordinate vector of a point within the shearing ball, relative to the volume-weighted center
\begin{equation}
    \mathbf{\Phi}_{\text{c}}\left(t\right) = V^{-1}\int_{V} \mathbf{\Phi}_{t}\left(\theta, \phi; r\right) \text{d}V ~.
\end{equation}

Since the polar points are always co-rotating with the center, the semi-major axis $\hat{e}_{\text{semi-major}} = \hat{e}_{z}$ and $b=r_{0}$.
The major axis evolves from $\alpha_{\text{major, 0}} \sim 45^\circ$ towards $\alpha_{\text{major, }\infty} \sim 0^\circ$, and necessarily $\alpha_{\text{minor}} = \alpha_{\text{major}} + 90^\circ$.

For the sake of a better intuition of the model, in Figs. \ref{fig:geometry_track_model} and \ref{fig:pitch_angles_model} we show the shape phase-space trajectories as well as the time evolution of the pitch angle for various shearing spheres with different $r_{0} / R_{0}$.
We show the time evolution over half an orbit.
The trajectories in shape phase-space differ only marginally between different sized spheres for $r_{0} \lesssim 0.1 \, R_{0}$.
However, at later times larger spheres tend to have slightly larger pitch angles and minor-to-major axis ratios. 

In Fig. \ref{fig:deformation_timescale_model} we show the dependence of the deformation timescale on the size of the sphere for $r_{0} < 0.1 \, R_{0}$.
We find a weak linear dependence on the size, which is approximately fit by
\begin{equation}
    t_{\text{deform}} \approx 0.065 \, t_{\text{orb}}\left(\bar{R}\right) ~, 
\end{equation}
where $\bar{R} = pR_{0} + (1-p)\left(R_{0} - r_{0}\right)$ is a characteristic radius lying between $R_{0} - r_{0}$ and $R_{0}$. Our linear fit yields $p\sim0.65$.

We do not show any results for $r_{0} > 0.1 \, R_{0}$ due to the large difference in orbital timescales, which leads to rapid deformation and even winding of the part of the sphere with $R \ll R_{0}$, while the rest has hardly moved.

\end{document}